\documentclass[11pt,a4paper]{article}
\usepackage{graphicx}
\usepackage{jcappub}
\usepackage{lscape}

\usepackage{amsmath, amssymb}
\usepackage[T1]{fontenc}
\usepackage[utf8]{inputenc}
\usepackage{lmodern}
\usepackage[english]{babel}

\usepackage[normalem]{ulem}

\newcommand{\simgt}{\lower.5ex\hbox{$\; \buildrel > \over \sim \;$}}
\newcommand{\simlt}{\lower.5ex\hbox{$\; \buildrel < \over \sim \;$}}

\makeatletter
\newsavebox{\@parc@ption}
\def\parcaption#1{%
\sbox{\@parc@ption}{\shortstack[l]{#1}}%
>\setbox\@tempboxa\hbox{\csname fnum@\@captype\endcsname}%
\@tempdima\columnwidth \advance\@tempdima-\wd\@tempboxa
\@tempdimb\@tempdima 
\ifdim\wd\@parc@ption>\@tempdimb \@tempdima\@tempdimb
\else\@tempdima\wd\@parc@ption\fi
\sbox{\@tempboxa}{\parbox[t]{\@tempdima}{#1}}%
\caption{\usebox{\@tempboxa}}}
\makeatother
\subheader{HUPD-1510, YITP-15-113}

\title{
Perturbed Newtonian description of the Lema\^{i}tre model with non-negligible pressure
}

\author{
Kazuhiro Yamamoto,${}^a$ Valerio Marra,${}^b$ \\
Viatcheslav Mukhanov,${}^c$  Misao Sasaki${}^d$
}

\affiliation{
$^a$Department of Physical Sciences, Hiroshima University, Higashi-hiroshima, Kagamiyama~1-3-1, 739-8526, Japan\\
$^b$Departamento de Física, Universidade Federal do Esp\'{\i}rito Santo, 
29075-910, Vit\'oria, ES, Brazil\\
$^c$Theoretical Physics, Ludwig Maxmillians University, Theresienstr.\ 37, 80333 Munich, Germany \\
$^d$Yukawa Institute for Theoretical Physics, Kyoto University, Kyoto 606-8502, Japan\\
}


\arxivnumber{1512.04240}

\abstract{
We study the validity of the Newtonian description of cosmological 
perturbations using the Lema\^{i}tre model, an \emph{exact} spherically symmetric solution of Einstein's equation. 
This problem has been investigated in the past for the case of a dust fluid. Here, we extend the previous 
analysis to the more general case of a fluid with non-negligible pressure, and, for the numerical examples, 
we consider the case of radiation ($P=\rho/3$).
We find that, even when the density contrast has a nonlinear amplitude, the Newtonian 
description of the cosmological perturbations using the gravitational 
potential $\psi$ and the curvature potential $\phi$ is valid as long as 
we consider sub-horizon inhomogeneities.
However, the relation $\psi+\phi={\cal O}(\phi^2)$ -- which holds for the case of a dust fluid -- 
is not valid for a relativistic fluid, and an effective anisotropic stress is generated.
This demonstrates the usefulness of the Lema\^{i}tre model which allows us to study in an exact nonlinear fashion the onset of anisotropic stress in fluids with non-negligible pressure.
We show that this happens when the characteristic scale of the 
inhomogeneity is smaller than the sound horizon and that the deviation is caused by the nonlinear effect of 
the fluid's fast motion. We also find that
$\psi+\phi= \max[{\cal O}(\phi^2),{\cal O}(c_s^2\phi \, \delta)]$
for an inhomogeneity with density contrast $\delta$ 
whose characteristic scale is smaller than the sound horizon, 
unless $w$ is close to $-1$, where $w$ and $c_s$ are 
the equation of state parameter and the sound speed of the fluid, respectively. 
On the other hand, we expect $\psi+\phi={\cal O}(\phi^2)$ to hold 
for an inhomogeneity whose characteristic 
scale is larger than the sound horizon, unless the amplitude 
of the inhomogeneity is  large and $w$ is close to $-1$. 
}

\begin{document}
\maketitle

\section{Introduction}
The Newtonian description of cosmological perturbations plays a key role in 
studies of structure formation as it often allows an intuitive understanding of the dynamics.
The availability of increasingly precise cosmological observations requires accurate
theoretical tools in order to describe the evolution of cosmological perturbations.
In this context, it might be useful to reconsider the properties of the perturbed 
Newtonian description of cosmological inhomogeneities.
In general, the Newtonian description of general relativity is based on the 
slow-motion approximation. Therefore, a system with non-negligible pressure
could be interesting in order to check the validity of the Newtonian description.

The Lema\^{i}tre model~\cite{Lemaitre:1933gd} is a spherically symmetric solution of Einstein's equation which can be used to obtain the exact dynamical evolution of a perfect fluid.
Therefore, the Lema\^{i}tre model provides us with the possibility of testing how well the exact solution can be reproduced in terms of the perturbed Newtonian description. 
This problem has been considered in Refs.~\cite{Biswas:2007gi,Paranjape:2008ai,VanAcoleyen:2008cy,Enqvist:2009hn} 
for the case of a dust-dominated universe (see also \cite{Rasanen:2010wz,Oliynyk:2014ufa,Sussman:2014wua}), which is usually named the Lema\^{i}tre-Tolman-Bondi (LTB) model~\cite{Lemaitre:1933gd,Tolman:1934za,Bondi:1947fta}.%
\footnote{These works were in part motivated by the ongoing debate on the backreaction proposal, according to which late-time matter inhomogeneities could affect the average expansion rate of the universe, possibly explaining away dark energy (see, e.g., the special focus issue~\cite{Andersson:2011za}). The present paper focuses on fluids with non-negligible pressure and, therefore, is not directly relevant to the backreaction proposal.}
Here, we aim at extending previous work to the more general case of a fluid with non-negligible pressure.

Contrary to its pressureless LTB limit, the Lema\^{i}tre model cannot be solved analytically.
Therefore, we integrate the relevant equations numerically as done, for instance, in~\cite{Bolejko:2005tk,Lasky:2006mg,Lasky:2006zz,Bolejko:2008ya,Alfedeel:2009ef,Lasky:2010vn,Marra:2011zp,Moradi:2013gf}.
We particularize our analysis to the case of radiation ($P=\rho/3$) and study the corresponding evolution for a set of different initial conditions.
We then study the validity of the description in terms of the perturbed Newtonian Friedmann-Lema\^{i}tre-Robertson-Walker (FLRW) metric by performing the coordinate transformation numerically for the set of different initial conditions previously considered.

This paper is organized as follows: In Section~\ref{model} we derive the basic equations 
for the Lema\^{i}tre model in a cosmological setup and in Section~\ref{dynamics}, for the case of a relativistic fluid, we solve numerically the equations and show the dynamics.
In Section~\ref{Newton}, we perform an exact coordinate transformation from the Lema\^{i}tre model to the perturbed FLRW metric.
Our results demonstrate the validity of the cosmological Newtonian description,
which assumes $|\psi|,~|\phi|\ll1$, where $\psi$ and $\phi$ are
the gravitational potential and the curvature potential in the Newtonian gauge, respectively. 
Then we show that $\phi+\psi={\cal O}(\phi^2)$ -- which holds in the case of a dust fluid -- 
is violated for a relativistic fluid with inhomogeneities at sub-sound horizon scales.
In Section~\ref{conform}, we show that the nonlinear effect of the fluid's fast motion is responsible for this failure, which is clarified 
by considering the second-order perturbations in spatially conformally 
flat spacetime, as a generalization of the perturbed FLRW metric.
This is an example of perturbations in a fluid with non-negligible pressure that effectively give rise to an anisotropic stress, which is an expected second-order effect~\cite[see, e.g.,][]{Bruni:1996im,Matarrese:1997ay,Hu:1998tj,Nakamura:2004wr,Tomita:2005et,Ballesteros:2011cm}.
Therefore, by studying the Lema\^{i}tre model we are able to follow the generation of anisotropic stress in an exact nonlinear way.
Section~\ref{conclusions} is devoted to summary and conclusions.
In Appendix~\ref{perturb} we develop the perturbation theory for the Lema\^{i}tre model.
Throughout this paper we adopt $c=1$, where $c$ is the velocity of light.

\section{The Lema\^{i}tre model}
\label{model}

In this Section we introduce the Lema\^{i}tre model~\cite[see, e.g.,][]{Alfedeel:2009ef} -- a spherically symmetric solution of Einstein's equation -- for the case of a perfect fluid with non-negligible pressure.%
\footnote{See, for example, Ref.~\cite{Grande:2011hm} for the case of a static fluid with anisotropic stress.}
Using comoving coordinates, it is customary to write the line element as:
\begin{eqnarray}
  ds^2=-e^{2\nu(t,r)}dt^2+e^{2\lambda(t,r)}dr^2+R^2(t,r) \, d\Omega^{2} \,,
\end{eqnarray}
where $d\Omega^2= d\theta^2+\sin^2\theta d\phi^2$ and $e^\nu$ is the lapse function, which cannot be gauged away when pressure gradients are present.
Einstein's equation gives then:
\begin{eqnarray}
G^{t}{}_{t}&=&\left(2{R''\over R}+{R'^2\over R^2}-2{R'\over R}\lambda'\right)e^{-2\lambda}
-\left({{\dot R}^2\over R^2}+2{\dot R\over R}\dot \lambda\right)e^{-2\nu}-{1\over R^2}
=-8\pi  G\rho \,,
\label{EEtt}
\\
G^{t}{}_{r}&=&\left({2{\dot R}'\over R}-{2\dot R\over R}\nu'-2{R'\over R}\dot\lambda
\right)e^{-2\nu}=0 \,,
\label{EEtr}
\\
G^{r}{}_{r}&=&\left({R'^2\over R^2}+{2R'\over R}\nu'\right)e^{-2\lambda}
-\left(2{{\ddot R}\over R}+{{\dot R}^2\over R^2}-2{\dot R\over R}\dot \nu\right)e^{-2\nu}
-{1\over R^2}=8\pi G P \,,
\label{EErr}
\\
G^{\theta}{}_{\theta}&=&\left({R''\over R}+{R'\over R}\nu'
+\nu''+\nu'^2-{R'\over R}\lambda'-\nu'\lambda'
\right)e^{-2\lambda} \,,
\nonumber\\
&&~~~~~~~~
+\left({\dot R\over R}\dot\nu-{\ddot R\over R}-\ddot\lambda
+\dot\lambda\dot\nu
-{\dot R\over R}\dot\lambda-\dot\lambda^2
\right)e^{-2\nu}=8\pi GP \,,
\label{EEhh}
\end{eqnarray}
where a dot and a prime denote differentiation with respect to $t$ and $r$, 
respectively, and $\rho$ and $P$ are the energy density and pressure of the
perfect fluid. 
With the use of (\ref{EEtr}),
equations~(\ref{EEtt}) and (\ref{EErr})  yield:
\begin{eqnarray}
M'(t,r)&=&4\pi G\rho R^2 R' \,,
\label{mtrr}
\\
\dot M(t,r)&=&-4\pi G P R^2 \dot R \,,
\label{mtrt}
\end{eqnarray}
respectively, where we defined the effective gravitating total mass $M$:
\begin{eqnarray}
  M(t,r)={R\over 2}\left({\dot R}^2 e^{-2\nu}-R'^2 e^{-2\lambda}+1
  \right).
\end{eqnarray}
As is clear from \eqref{mtrr}, $M$ is related to the local density $\rho$ through the Euclidean volume element. Consequently, $M$ does not coincide with the invariant mass.  We call $M$ the effective gravitating mass because it enters the generalized Friedmann equation~\eqref{aeq}. See Refs.~\cite{1970JMP....11.1382C,1970JMP....11.1392C, Alfedeel:2009ef,Marra:2011zp} for more details.

In the comoving coordinate we are using, the four-velocity of the perfect fluid is $v^\mu=(e^{-\nu},0,0,0)$. The equations of motion of the
perfect fluid are then: 
\begin{eqnarray}
&&{\nu'}+{P'\over \rho+P}=0 \,,
\label{Trr}
\\
&&{\dot\lambda}+{\dot\rho\over \rho+P}+2{\dot R\over R}=0 \,,
\label{Ttt}
\end{eqnarray}
which are integrated as:
\begin{eqnarray}
 \nu({t,r})&=&\nu_0(t)-\int_{r_0}^{r}{d \bar r P'\over \rho+P}=\nu_0(t)-{w\over 1+w}\ln {\rho(t,r)\over \rho(t,r_0)} \,,
\label{sigmasol}
\\
\lambda({t,r})&=&\lambda_i(r)-\int_{t_i}^{t}{d \bar t \dot\rho\over \rho+P}-2\ln
{R(t,r)\over R_i(r)}=\lambda_i(r)-{1\over 1+w}\ln {\rho(t,r)\over \rho_i(r)}
-2\ln {R(t,r)\over R_i(r)} ,
\label{lambdasol}
\end{eqnarray}
where in the last step the equation of state $P=w\rho$ with a constant parameter $w$ was assumed.
We assume that the sound speed is $c_s=\sqrt{w}$, unless explicitly noted otherwise.
In the above equations, $\rho_i(r)=\rho(t_i,r)$, $R_i(r)=R(t_i,r)$ and $\lambda_i(r)=\lambda(t_i,r)$ are evaluated at the initial time $t_i$, and $\nu_0(t)=\nu(t,r_0)$ is an arbitrary function which is determined once we fix the residual temporal gauge.

By combining (\ref{Trr}), (\ref{Ttt}) and (\ref{EEtr}) we obtain
\begin{eqnarray}
\dot\rho+P'{\dot R\over R'}+(\rho+P)\left[
{{\dot R}'\over R'}+2{\dot R\over R}\right]=0 \,, 
\label{EMe}
\end{eqnarray}
which with (\ref{Ttt}) yields
\begin{eqnarray}
\dot\lambda={1\over R'}\left(
{P'\dot R\over \rho+P}+{\dot R}'\right). 
\label{EMf}
\end{eqnarray}
Then by integrating (\ref{EEtr}) one obtains the radial scale factor $e^\lambda$ as a function of the angular scale factor $R$ and the lapse function $\nu$:
\begin{eqnarray}
e^{\lambda}={R'\over \sqrt{1+2E(r)}}\exp\left(-\int_{t_i}^t d \bar t \, {\nu'\dot R
\over R'}\right),
\label{sigmatrpg2}
\end{eqnarray}
where $E(r)$ is related to 
$\lambda_i(r)$ by $e^{\lambda_i(r)}=R'_i(r)/\sqrt{1+2E(r)}$.

Finally, the combination $G^\theta_{\theta}-G^r_r$ from (\ref{EErr}) and (\ref{EEhh}), which have not been used so far as a consequence of the Bianchi identity, gives:
\begin{eqnarray}
&& \left({R''\over R}-{R'^2\over R^2}-{R'\over R}\nu'
+\nu''+\nu'^2-{R'\over R}\lambda'-\nu'\lambda'
\right)e^{-2\lambda}
\nonumber\\
&&~~~~~~~~~~~~
+\left({\ddot R\over R}+{{\dot R}^2\over R^2}
-{\dot R\over R}\dot\nu
-\ddot\lambda-\dot\lambda^2
-{\dot R\over R}\dot\lambda
+\dot\nu\dot\lambda
\right)e^{-2\nu}+{1\over R^2}=0 \,.
\label{gmg}
\end{eqnarray}
We will use this equation in order to check our numerical computations (see text below~(\ref{tracefree})).

\subsection{Redefinition of scale-factor functions}

We will find it useful to redefine the functions $\lambda$ and $R$ that describe the radial and angular scale factors by means of the functions $a(t,r)$ and ${\cal E}(t,r)$ according to:
\begin{eqnarray}
 R(t,r)=ra(t,r),~~~~~e^{\lambda(t,r)}=a(t,r)e^{{\cal E}(t,r)} \,.
\end{eqnarray}
Thanks to this redefinition the line element now reads:
\begin{eqnarray}
ds^2=-e^{2\nu(t,r)}dt^2+a^2(t,r)e^{2{\cal E}(t,r)}dr^2+a^2(t,r)r^2 d\Omega^{2} \,.
\label{GTBGTB}
\end{eqnarray}
To our knowledge, this parametrization has not been used before.

Equations (\ref{mtrr}) and (\ref{mtrt}) become:
\begin{eqnarray}
M'(t,r)&=&4\pi G\rho R^2 R'=4\pi G\rho \, r^2a^2(t,r)(a(t,r)+ra'(t,r)),
\label{mtrr2}
\\
\dot M(t,r) &=& -4\pi G P R^2 \dot R=-4\pi GP \, r^3a^2(t,r)\dot a(t,r),
\label{mtrt2}
\end{eqnarray}
where the effective gravitating mass is now:
\begin{eqnarray}
M(t,r)=
{ra(t,r)\over 2}\left[e^{-2\nu}r^2 \dot a^2(t,r)
-{e^{-2{\cal E}(t,r)}\left(1+{ra'(t,r)\over a(t,r)}\right)^2}
+1
\right].
\label{defM}
\end{eqnarray}
From the previous equation one obtains:
\begin{eqnarray}
&&\dot a(t,r)=e^{\nu(t,r)}\sqrt{{2M(t,r)\over r^3a(t,r)}+{e^{-2{\cal E}(t,r)}\over r^2}
\left[1+{ra'(t,r)\over a(t,r)}\right ]^2-{1\over r^2}},
\label{aeq}
\end{eqnarray}
which can be used for the time evolution of $a(t,r)$.
Equation (\ref{lambdasol}) is rephrased as
\begin{equation}
{\cal E}({t,r})={\cal E}_i(r)-\int_{t_i}^{t}{d \bar t \dot\rho\over \rho+P}-3\ln{a(t,r)\over a_i(r)}
= {\cal E}_i(r)-{1\over 1+w}\ln {\rho(t,r)\over \rho_i(r)}
-3\ln {a(t,r)\over a_i(r)} ,
\label{cale2}
\end{equation}
where $a_i(r)=a(t_i,r)$ and ${\cal E}_i(r)=\lambda_i(r) - \ln  a_i(r)$.
Equations (\ref{EMe})--(\ref{sigmatrpg2}) are rewritten as:
\begin{eqnarray}
\dot\rho &=& - {rP'\dot a(t,r)\over a(t,r)+ra'(t,r)}-(\rho+P)\left[
{\dot a+r\dot a'\over a+ra'}+2{\dot a\over a}\right] \,,
\label{aEMe} \\
\dot {\cal E} &=& {r\over a(a+ra')}\left(-\dot a a'+a{\dot a}'
+a\dot a{P'\over \rho+P}\right) \,,
\label{EMfz} \\
e^{\cal E} &=& {1+{ra'/ a}\over \sqrt{1+2E(r)}}\exp
\left(-\int_{t_i}^t d \bar t{r\nu'\dot a/a \over 1+ra'/a}\right) \,.
\label{eeee}
\end{eqnarray}

In summary, the unknown functions $a(t,r)$, $M(t,r)$, ${\cal E}(t,r)$ and $\rho(t,r)$ are determined using equations \eqref{aeq}, (\ref{mtrt2}), (\ref{EMfz}),\footnote{Equivalently, one could use \eqref{cale2} or \eqref{eeee}.} (\ref{aEMe}), respectively, while equation~(\ref{sigmasol}) is used to determine $\nu(t,r)$ at each time step.

\section{Dynamical evolution}
\label{dynamics}

\subsection{Initial and boundary conditions}

We will now discuss the initial conditions necessary in order to integrate numerically~\cite[see, e.g.,][]{Alfedeel:2009ef,Moradi:2013gf} the system of coupled differential equations presented in the previous section.

We adopt a gaussian initial density contrast:
\begin{eqnarray}
\rho_i(r)=\bar \rho_i \left [1+A\exp\left(-{r^2\over L^2}\right)\right ] \,,
\label{rhozero}
\end{eqnarray}
where the parameter $A$ specifies the amplitude of the initial inhomogeneity,
$L$ is its comoving characteristic size, $\bar \rho_i = 3 H_i^2/8\pi G$ is the background (critical) density, $H=\dot {\bar a}/ \bar a$ is the background Hubble parameter, and $\bar a(t)=a(t,r=\infty) $ is the background scale factor.%
\footnote{An overbar on $\rho$ and $a$ indicates the corresponding background quantities.}
As a consequence of the chosen profile, the background FLRW model is recovered only at infinity -- that is, $\bar \rho_i =\rho_i(t_i,r=\infty)$ -- rather than at a finite radius~\cite[see][]{Valkenburg:2012td,Valkenburg:2013qwa}.
Also, as is clear from \eqref{eeee}, if $E(r=\infty)=0$, then the background model is spatially flat.
We adopt $t_i=0$ and time is given in units of the Hubble time $H_i^{-1}$, while the radial coordinate $r$ is given in units of $L$.
Furthermore, the dimensionless parameter $\ell=L  H_i$
specifies the characteristic scale of the spherical overdensity in units
of the initial Hubble radius.
In the present paper, we consider an initial overdensity, that is, $A>0$.

To better understand the dynamical evolution and also in connection with the discussion of Section~\ref{conform}, it is useful to calculate the comoving sound horizon scale, which we estimate as: 
\begin{eqnarray}
r_s={c_s \over \bar a(t) H(t)}={w^{1/2}L\over \ell} \; {\bar a(t)^{(1+3w)/2}
\over (1+3w)/2} \,,
\label{shor}
\end{eqnarray}
where the background solution ${\bar a}(t)\propto t^{2/3(1+w)}$ was used.
Then we adopt $\nu_0(t)=\nu(t,r_0)=0$, where we set $r_0=\infty$.
Thanks to this choice, when $w=0$ the metric reduces to the LTB metric in synchronous gauge.
However, in the presence of the pressure the lapse function cannot be gauged away and is given by \eqref{sigmasol}:
\begin{eqnarray}
&&\nu(t,r)=-{w\over 1+w}\ln {\rho(t,r)\over \bar \rho(t)} \,.
\label{sigmasolZ}
\end{eqnarray}
We then fix the spatial gauge freedom left with $a_i(r)=1$.
The initial condition for the gravitating mass is then from \eqref{mtrr2}:
\begin{equation}
M_i(r)=4\pi G\int_0^r d \bar r \bar r^2\rho_i(\bar r)
= \bar M_i \left (1 +  \frac{I_{\delta_i}}{4 \pi  r^3 /3} \right)  \,,
\end{equation}
where $\bar M_i = 4 \pi G \bar \rho_i r^3 /3$ and $I_{\delta_i}=4\pi \int_0^r d \bar r \bar r^2\delta_i(\bar r)$, and, therefore, the last term in the previous equation is the average integrated density contrast.
From \eqref{rhozero} it is $\delta_i(r)=A e^{-r^2/L^2}$, where as always it is $\delta= \rho /\bar \rho -1$.

Regarding ${\cal E}(t,r)$, from \eqref{eeee} we obtain:
\begin{equation}
e^{{\cal E}_i(r)} = {1\over \sqrt{1+2E(r)}} \,,
\label{eerel}
\end{equation}
that is, the function $E(r)$ gives the initial spatial curvature, as it is clear from \eqref{GTBGTB}.
We will consider two different choices for the curvature function $E(r)$ and therefore for ${\cal E}_i(r)$.

\subsubsection{Zero initial spatial curvature}
\label{zero}

The first case which we will consider is of zero initial spatial curvature, that is:
\begin{equation}
E(r)=0~~~~(\text{and consequently }{\cal E}_i(r)=0) \,.
\end{equation}
An initial condition for ${\cal E}_i(r)$ is equivalent to an initial condition for $\dot a_i (r)$, that is, the initial velocity of the various shells.
From \eqref{aeq}  one finds:
\begin{equation}
e^{-2\nu_i(r)} \dot a_i(r)^2= 2 M(t_i,r)/r^3 = \frac{8 \pi G}{3} \bar \rho_i
 \left ( 1  + \frac{I_{\delta_i}}{4 \pi r^3/3} \right )  \,,
\end{equation}
that is an initial velocity which is proportional to the initial integrated overdensity.
This initial condition appears to be dominated by decaying modes as overdensities are expanding away faster than the background.
In the previous equation $e^{-\nu_i(r)} \dot a_i(r) = da/d \tau$ is the derivative of $a_i(r)$ with respect to the proper time of the comoving observer/fluid as $d/d\tau =v^{\mu} \, \partial_{\mu}= e^{-\nu} \partial/\partial t$.

\begin{figure}[t]
\begin{center}
    \hspace{0mm}\scalebox{0.5}{\includegraphics{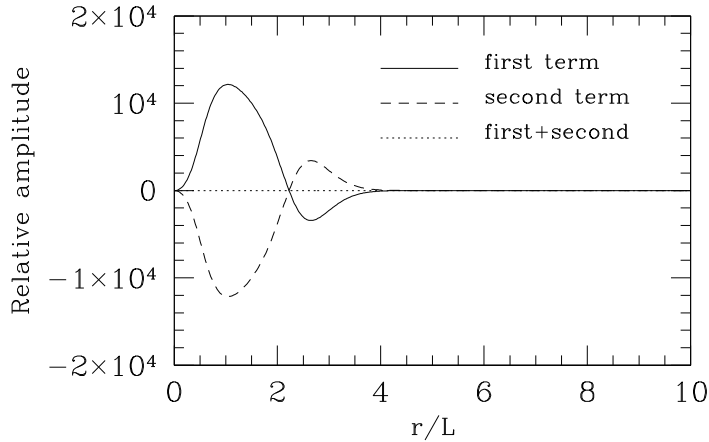}}
\vspace{0cm}
\caption{A snapshot to demonstrate the validity of our numerical results.
 The solid curve is the first term of the extra equation (\ref{tracefree}), while the
 dashed curve is the second term, as functions
 of $r$ in units of the inhomogeneity scale $L$ (see equation (\ref{rhozero})).
 The parameters relative to this snapshot are the same as those of the right 
 panels of Figure~\ref{fig:w03NLETLT}.
  \label{fig:gmg}}
\end{center}
\end{figure}

\subsubsection{Uniform initial expansion}
\label{uniform}

The second case we will consider is of initial uniform expansion, that is:
\begin{eqnarray}
e^{-\nu_i(r)} \dot a_i(r) =  H_i \,,
\label{ICA2}
\end{eqnarray} 
which implies:
\begin{align}
E(r) &= \frac{1}{2}   H(t_i)^2 r^2 - \frac{M_i(r)}{r} 
= \frac{1}{2}r^2 \left[  H(t_i)^2- \frac{8 \pi G}{3} \bar \rho_i
 \left ( 1  + \frac{I_{\delta_i}}{4 \pi r^3/3} \right )  \right] \,,
\end{align}
and ${\cal E}_i(r)$ is given by \eqref{eerel}.
Clearly, it is $E(r=\infty)=0$.
This initial condition features a mixture of growing and decaying modes.
The dynamical evolution relative to these two initial conditions was studied by 
Ref.~\cite{Marra:2007pm} in the case of the pressureless LTB model.

\subsection{Numerical Analysis}
\label{nume}

\begin{figure}[t]
\begin{center}
\scalebox{0.65}{
    \hspace{-35mm}
\includegraphics{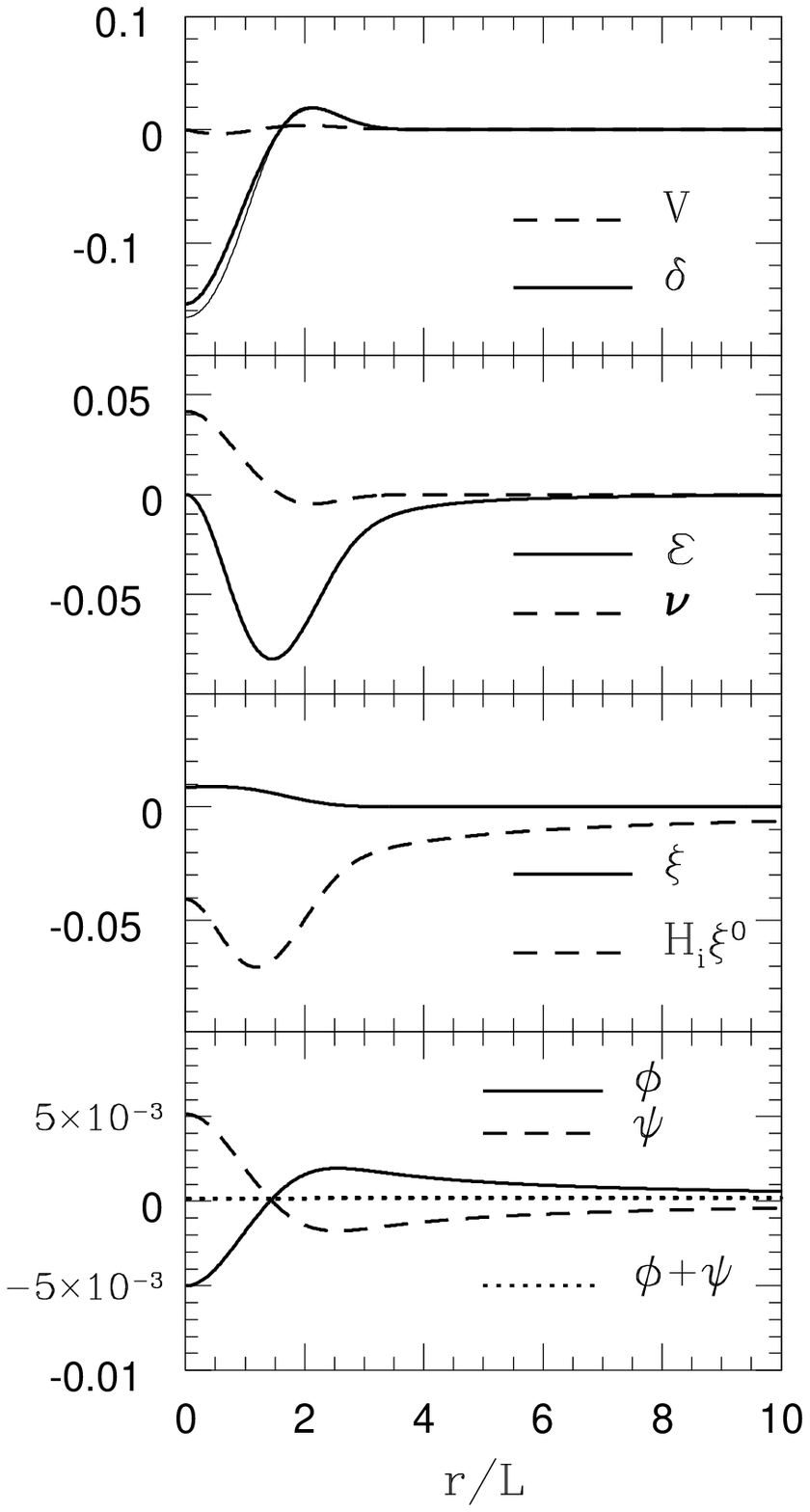}
    \hspace{-90mm}
\includegraphics{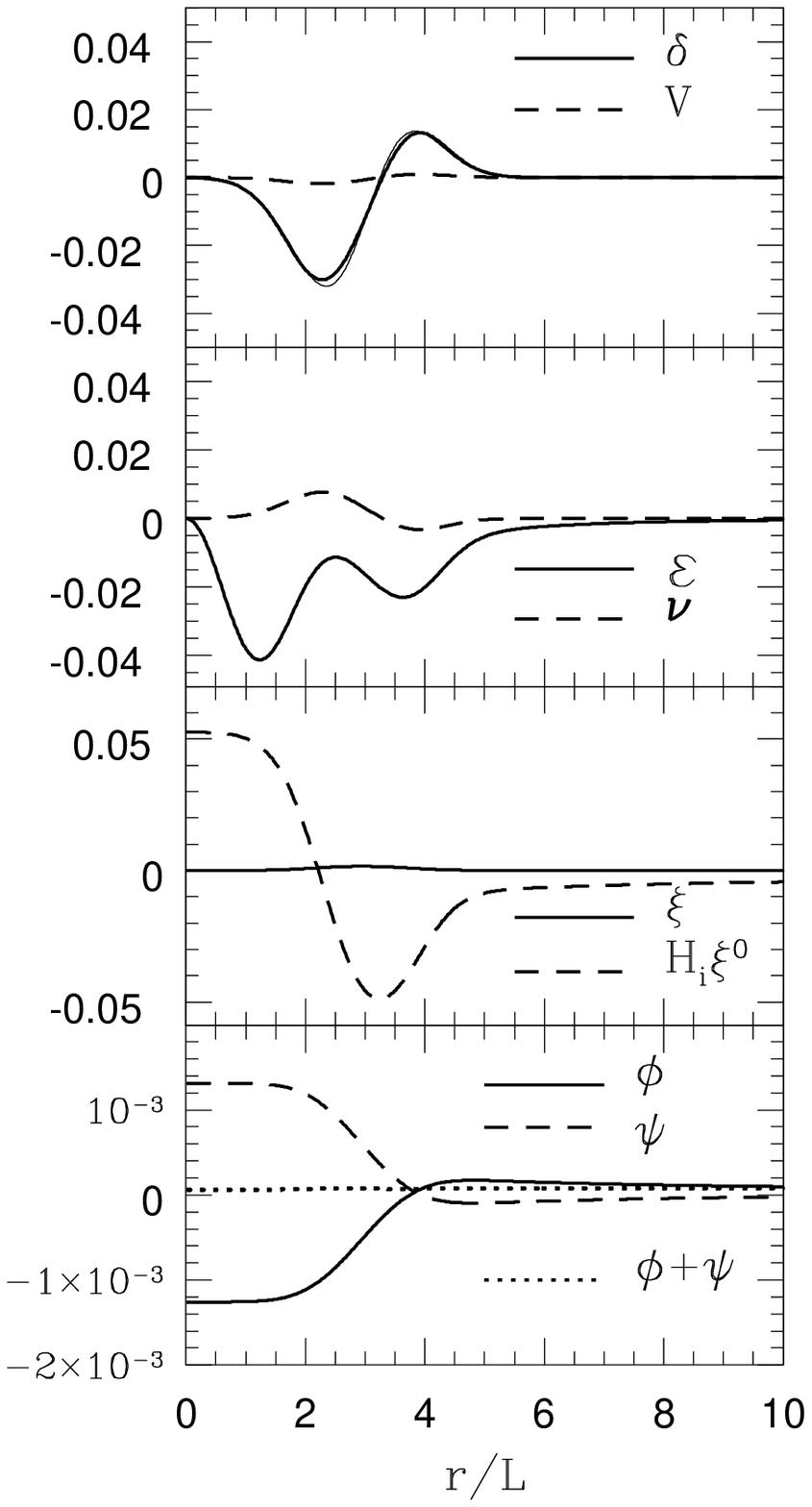}}
\caption{Numerical results for the evolution of 
a relativistic fluid ($w=1/3$) at the times
$H_it=5~(\bar a=3.3)$ (left panels) and 
$H_it=20~(\bar a=6.4)$ (right panels) for an initial density contrast of $A=0.3$
and characteristic scale of one initial Hubble radius $\ell=LH_i=1$.
The initial spatial curvature is set to zero as explained in Section~\ref{zero}.
In the top panels, the thick solid curve is the contrast $\delta(r)$, the dashed curve is the velocity $V$
defined in (\ref{defV}), and the thin solid curve is the density contrast in the
linear theory (see the Appendix~\ref{perturb}).
See Figure~\ref{fig:w03L1timeevolution} (left panels) for the time evolution of the density contrast. 
The second panels from the top show the corresponding
profiles of the lapse function $\nu(r)$ and ${\cal E}(r)$.
The details of the lower two panels are explained in Section~\ref{Newton}. 
The third panels from the top show $\xi(r)$ and $H_i\xi^0(r)$,
while the bottom panels show $\phi(r)$, $\psi(r)$,
and $\phi(r)+\psi(r)$. In these two panels, 
the curves are the results of the full nonlinear transformation defined by~(\ref{eqa}-\ref{eqb}).
See Section~\ref{nume} for details.\label{fig:w03L1ETLT}}
\end{center}
\end{figure}

\begin{figure}[t]
\begin{center}
    \hspace{0mm}\scalebox{0.65}
    {\hspace{-35mm}\includegraphics{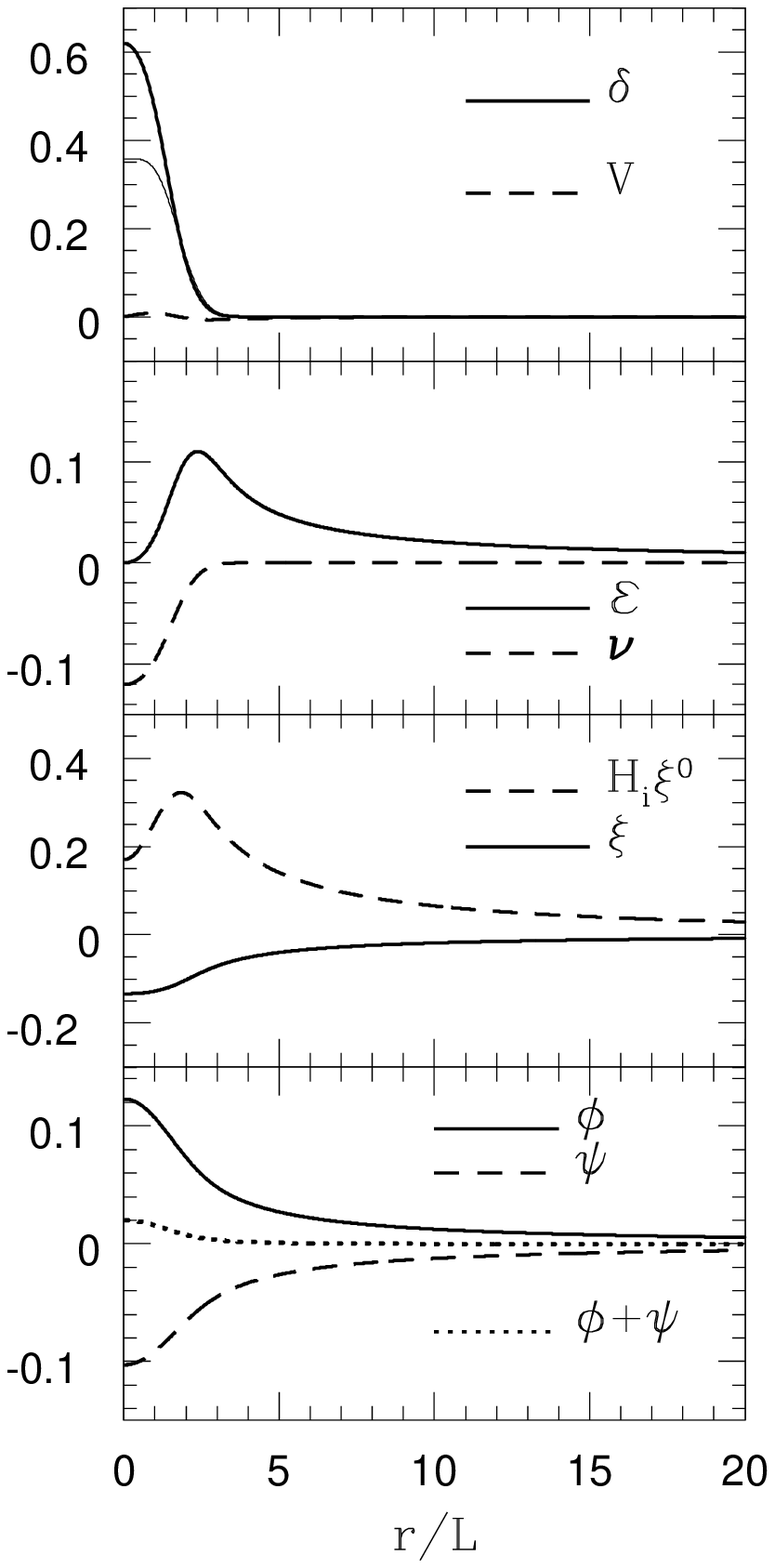}
     \hspace{-90mm}\includegraphics{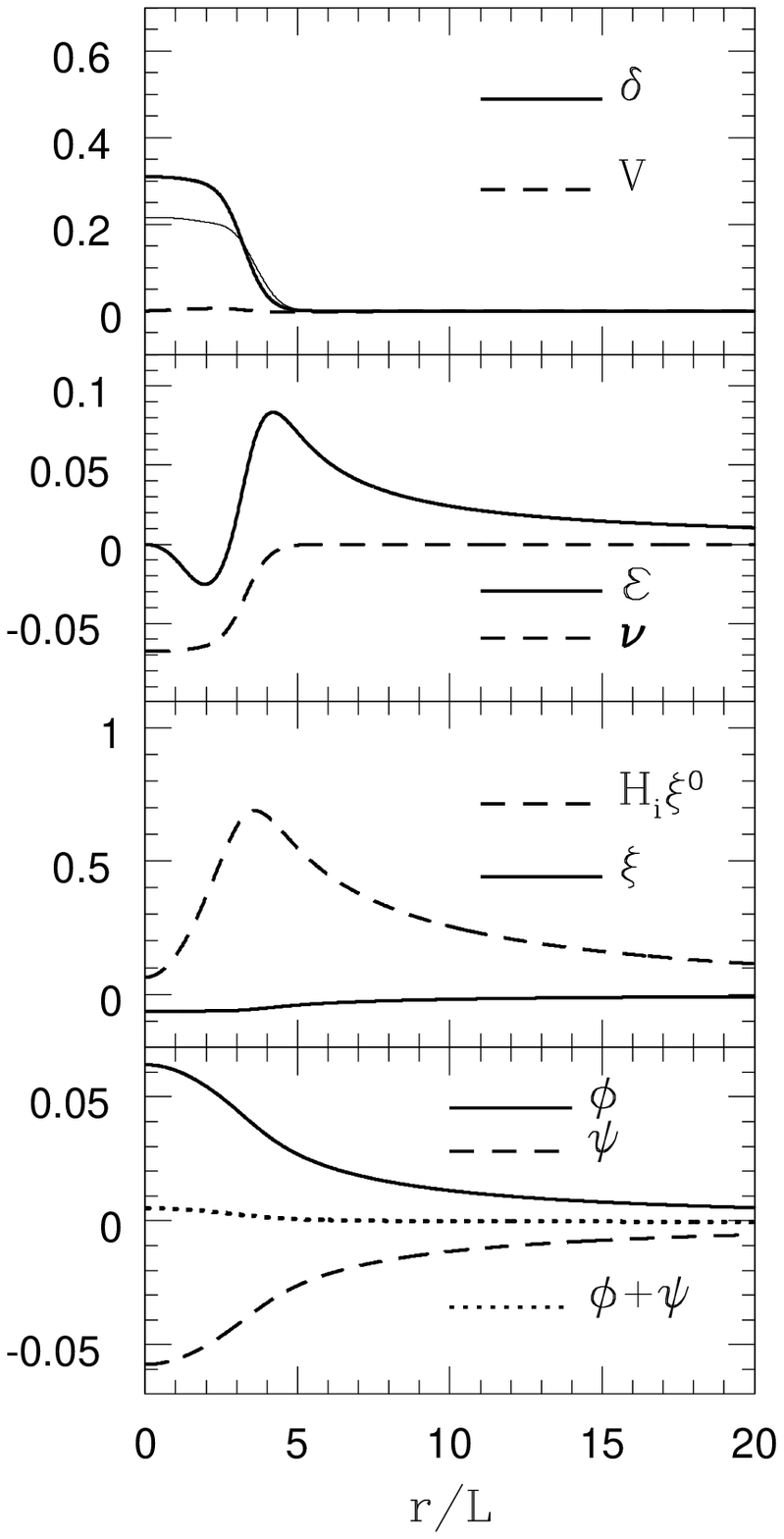}}
\caption{Same as in Figure~\ref{fig:w03L1ETLT}, but in the case of the uniform initial expansion condition described 
in Section~\ref{uniform}, at the times $H_it=5~(\bar a=3.3)$ (left panels) and $H_it=20~(\bar a=6.4)$ (right panels). See Figure~\ref{fig:w03L1timeevolution} (right panels) for the time evolution of the density contrast. 
  \label{fig:w03LNETLTIC}}
\end{center}
\end{figure}

\begin{figure}[t]
\begin{center}
\scalebox{0.65}{
    \hspace{-35mm}
\includegraphics{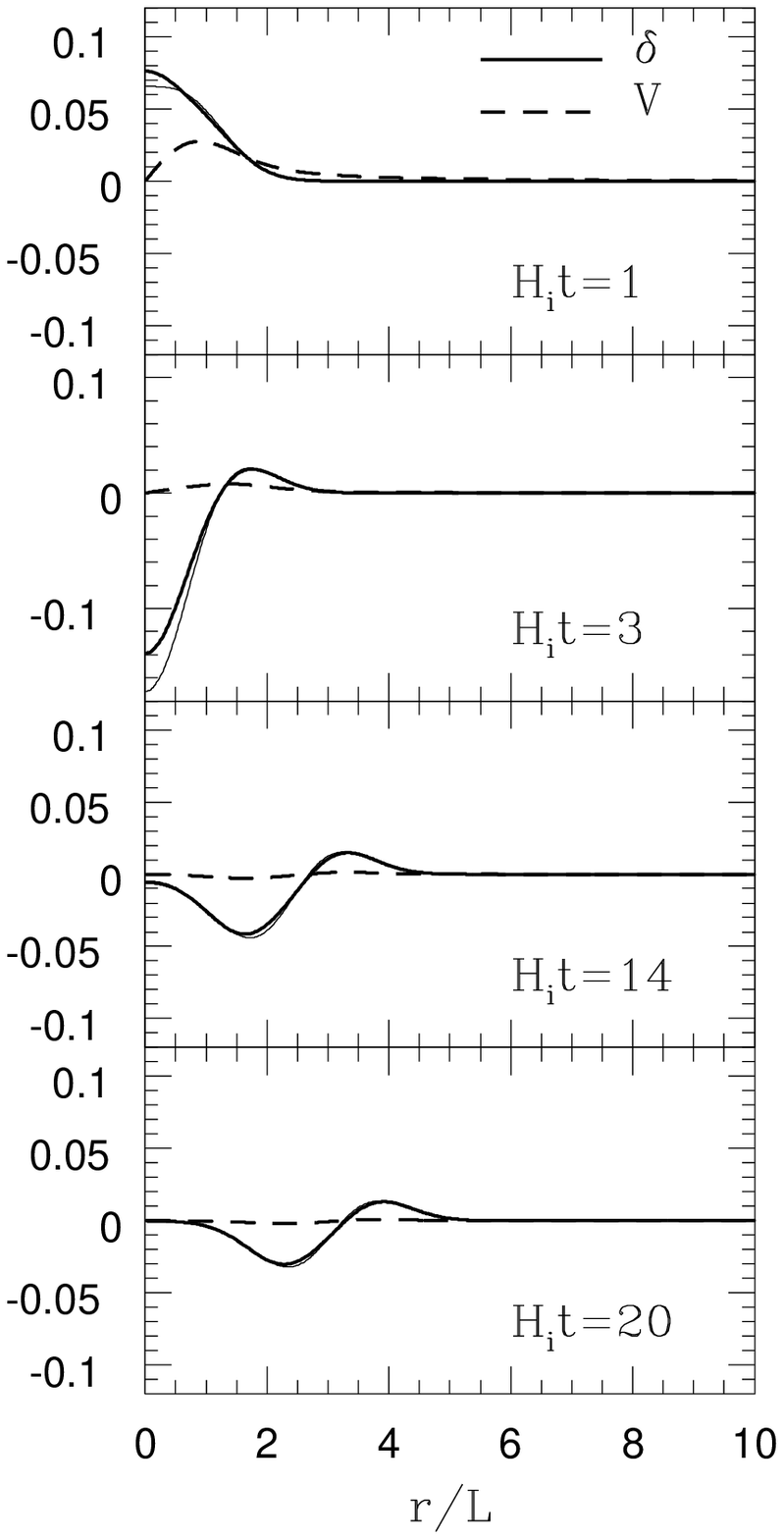}
    \hspace{-90mm}
\includegraphics{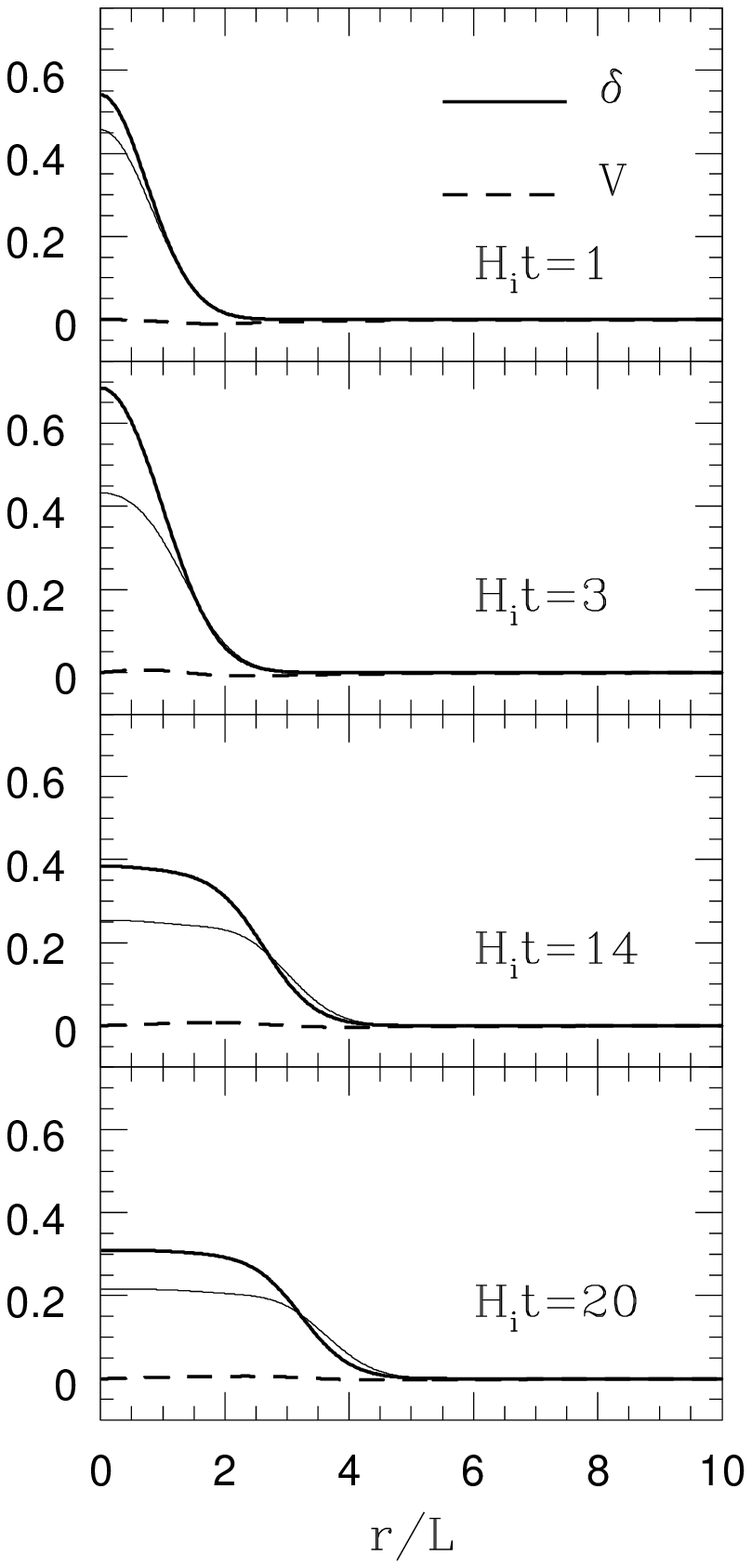}}
\caption{
Time evolution of the density contrast relative to Figure~\ref{fig:w03L1ETLT} (left panels) and Figure~\ref{fig:w03LNETLTIC} (right panels).
The left panels adopt the zero-curvature initial condition discussed in 
Section~\ref{zero}, while the right panels adopt the uniform initial expansion condition 
discussed in Section~\ref{uniform}.
\label{fig:w03L1timeevolution}}
\end{center}
\end{figure}

\begin{figure}[t]
\begin{center}
    \hspace{0mm}\scalebox{0.65}
    {\hspace{-35mm}\includegraphics{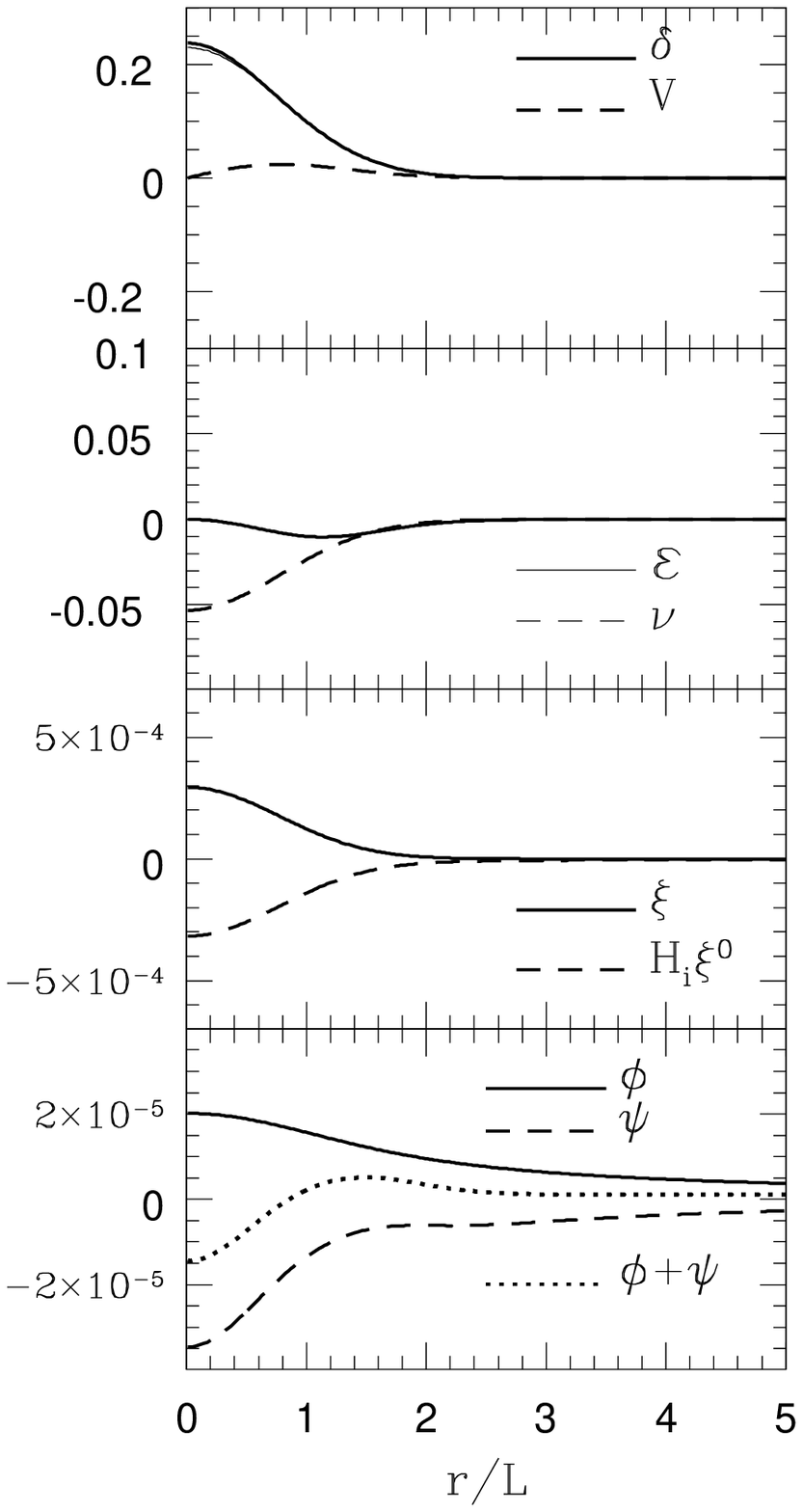}
     \hspace{-90mm}\includegraphics{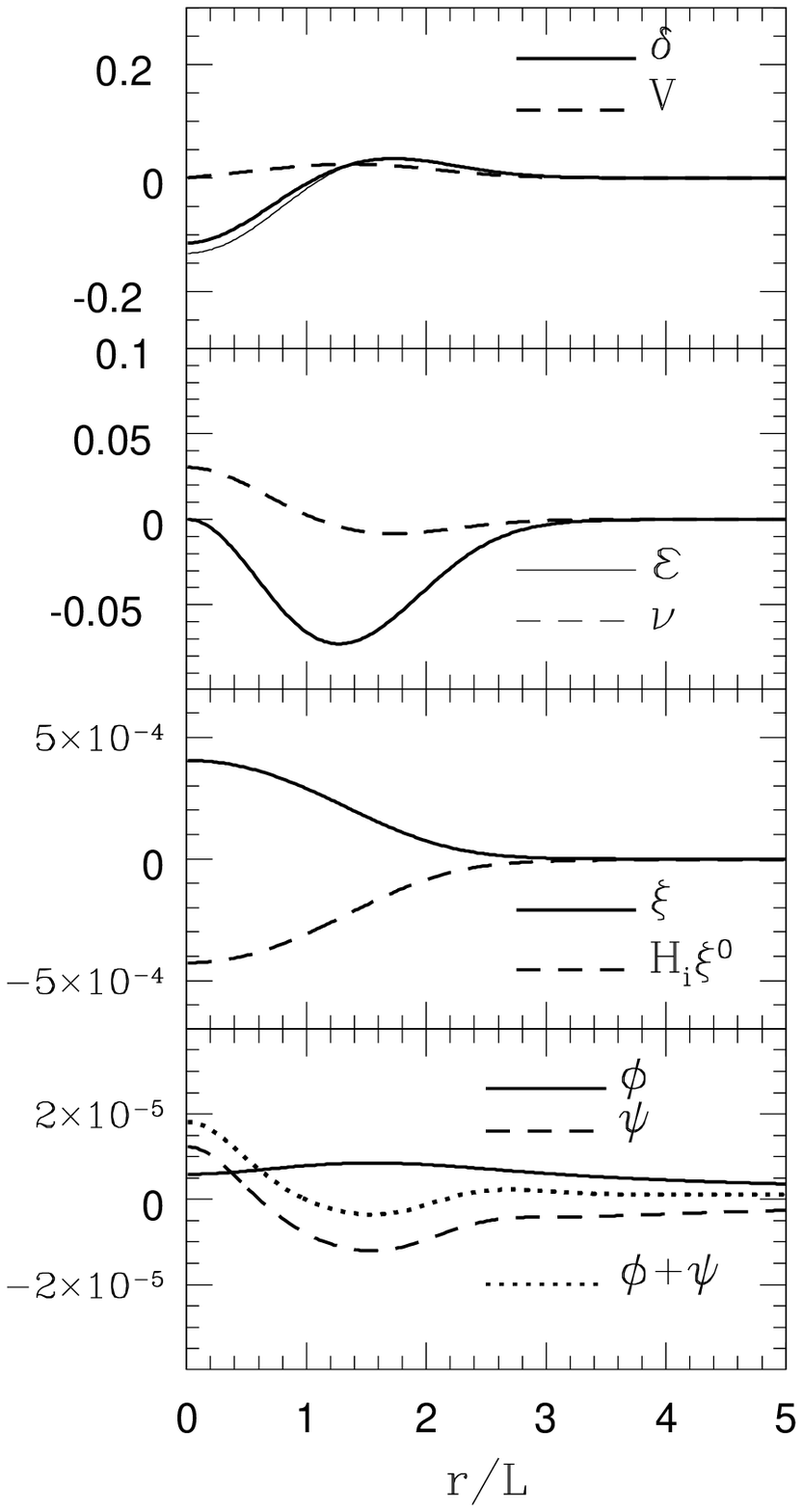}}
\caption{Same as Figure \ref{fig:w03L1ETLT}, but for $A=0.3$ and $\ell=0.01$
at the time $H_it=0.005$ (left panels) and $H_it=0.02$ (right panels).
  \label{fig:w03A03ETLT}}
\end{center}
\end{figure}

\begin{figure}[t]
\begin{center}
    \hspace{0mm}\scalebox{0.65}
    {\hspace{-35mm}\includegraphics{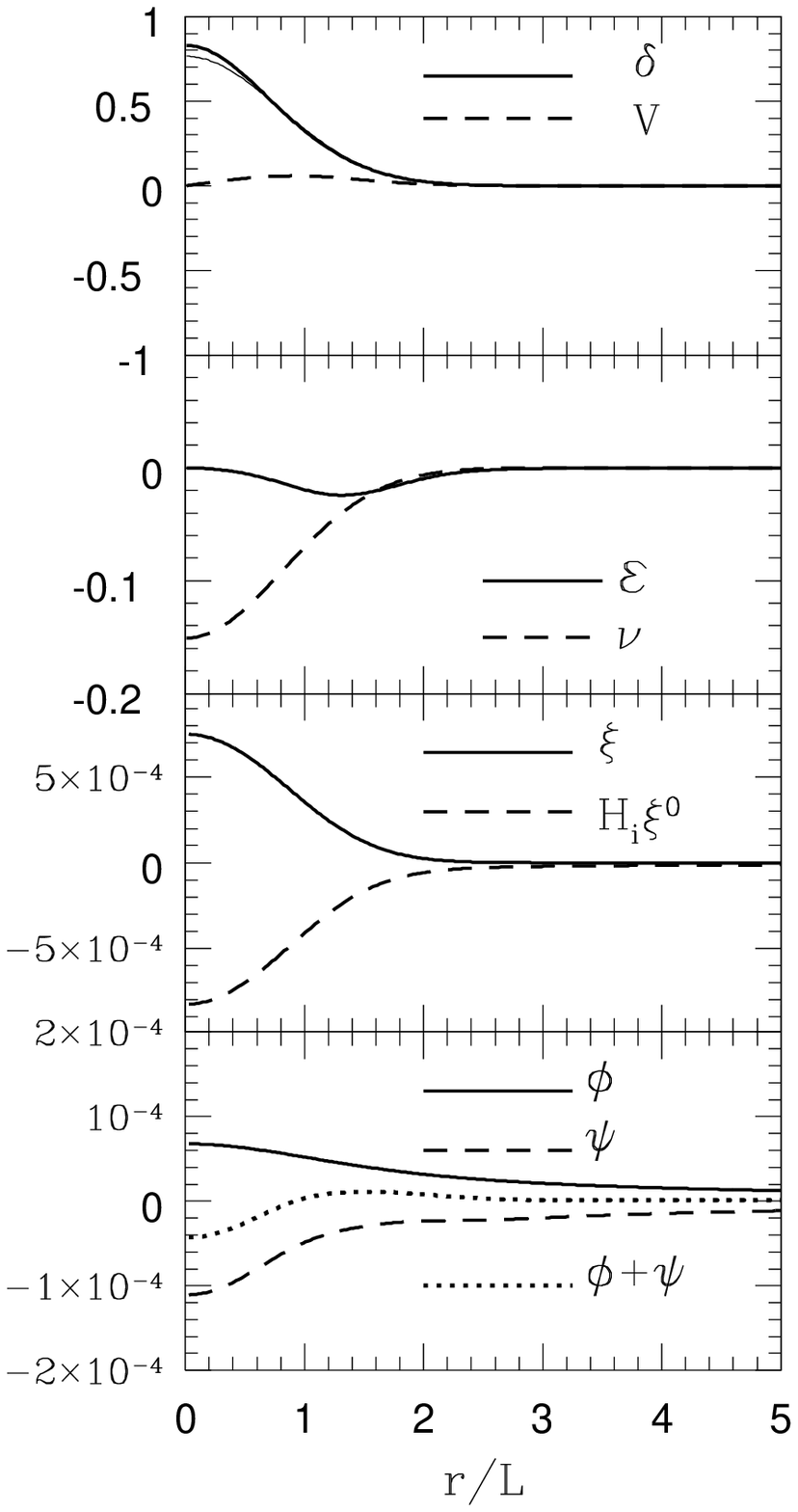}
     \hspace{-90mm}\includegraphics{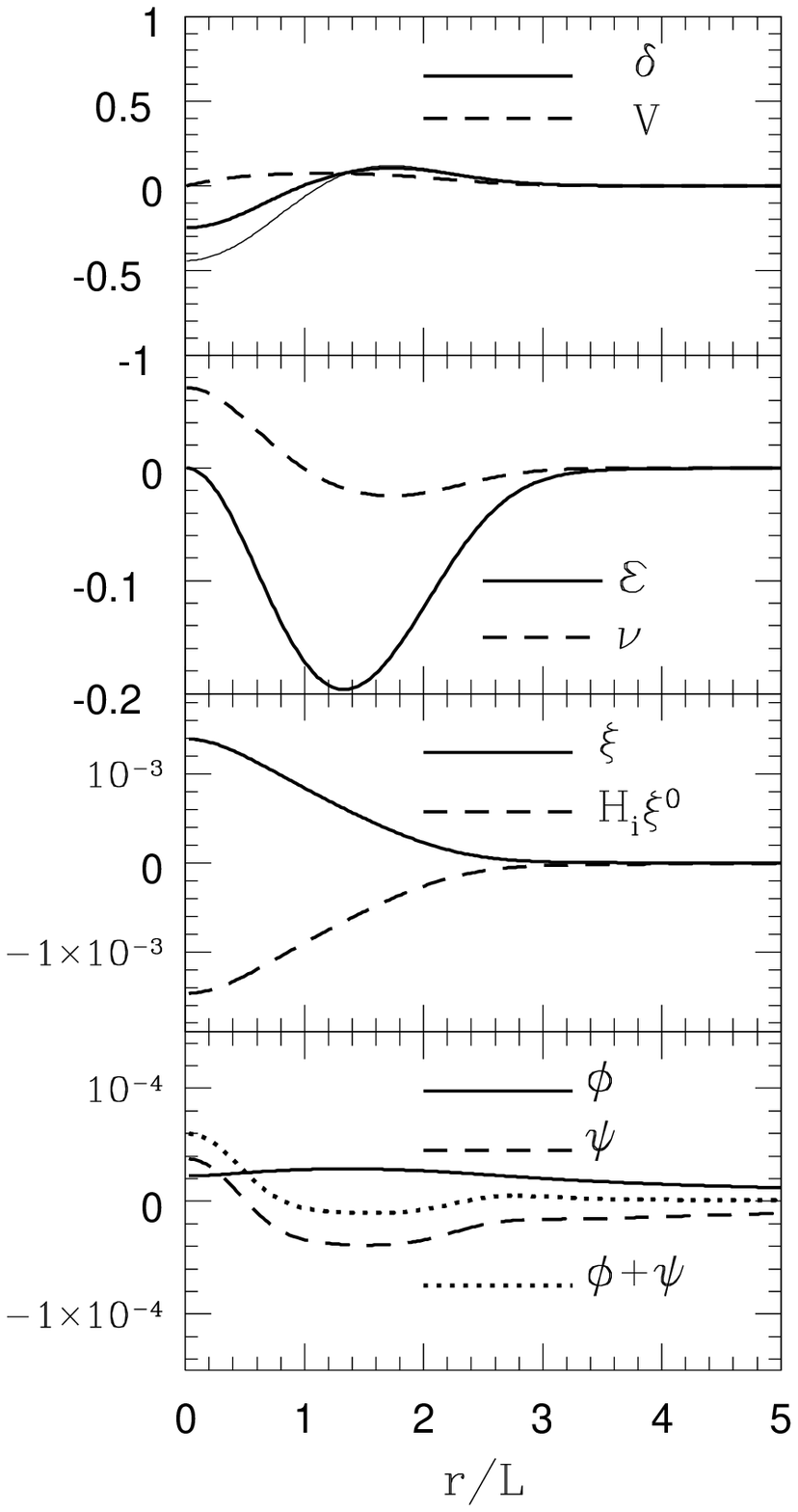}}
\caption{Same as Figure \ref{fig:w03L1ETLT}, but for $A=1$ and $\ell=0.01$ 
at the time $H_it=0.005$ (left panels) and $H_it=0.02$ (right panels).
  \label{fig:w03NLETLT}}
\end{center}
\end{figure}

\begin{figure}[t]
\begin{center}
\scalebox{0.65}{
    \hspace{-35mm}
\includegraphics{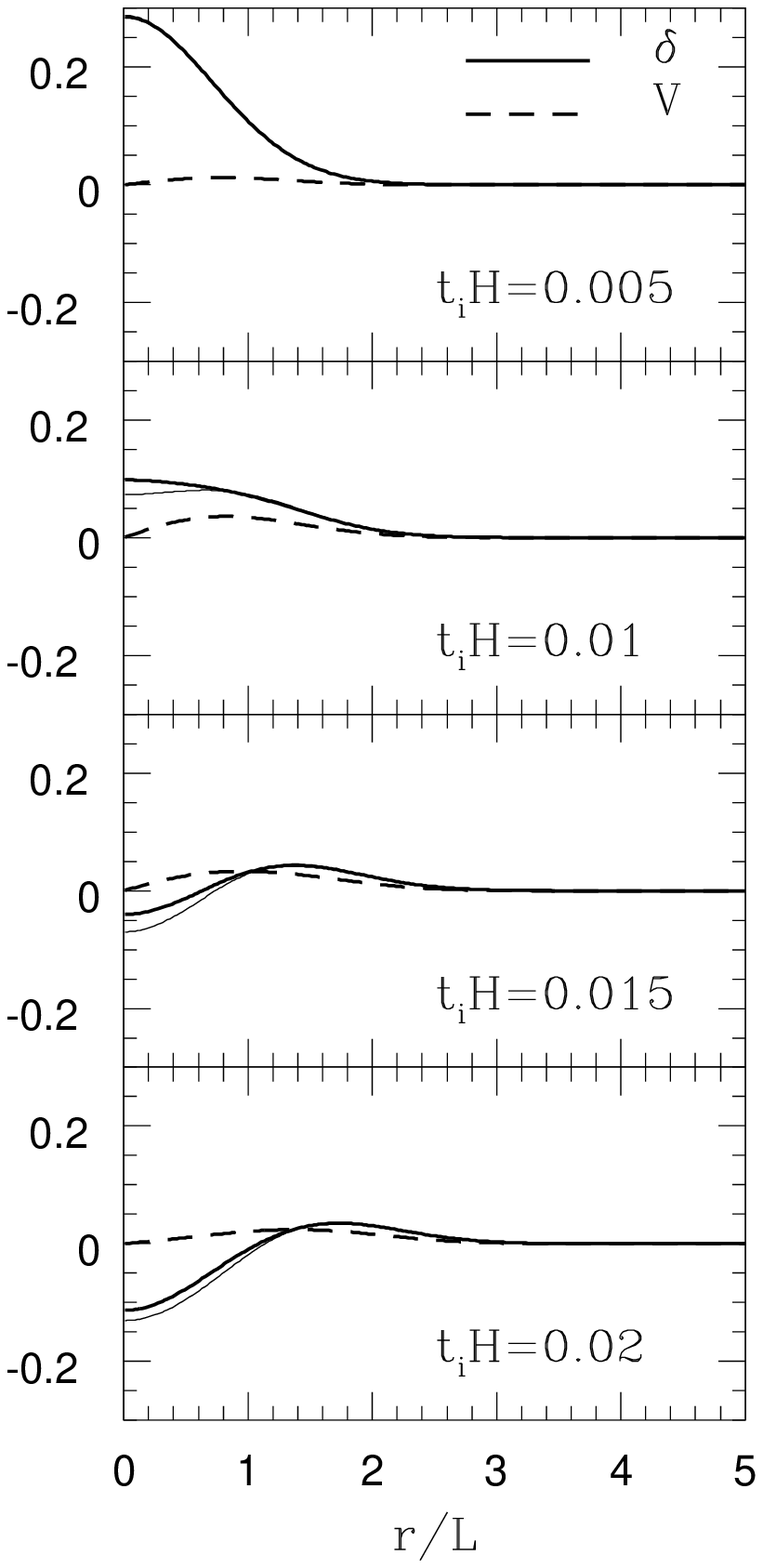}
    \hspace{-90mm}
\includegraphics{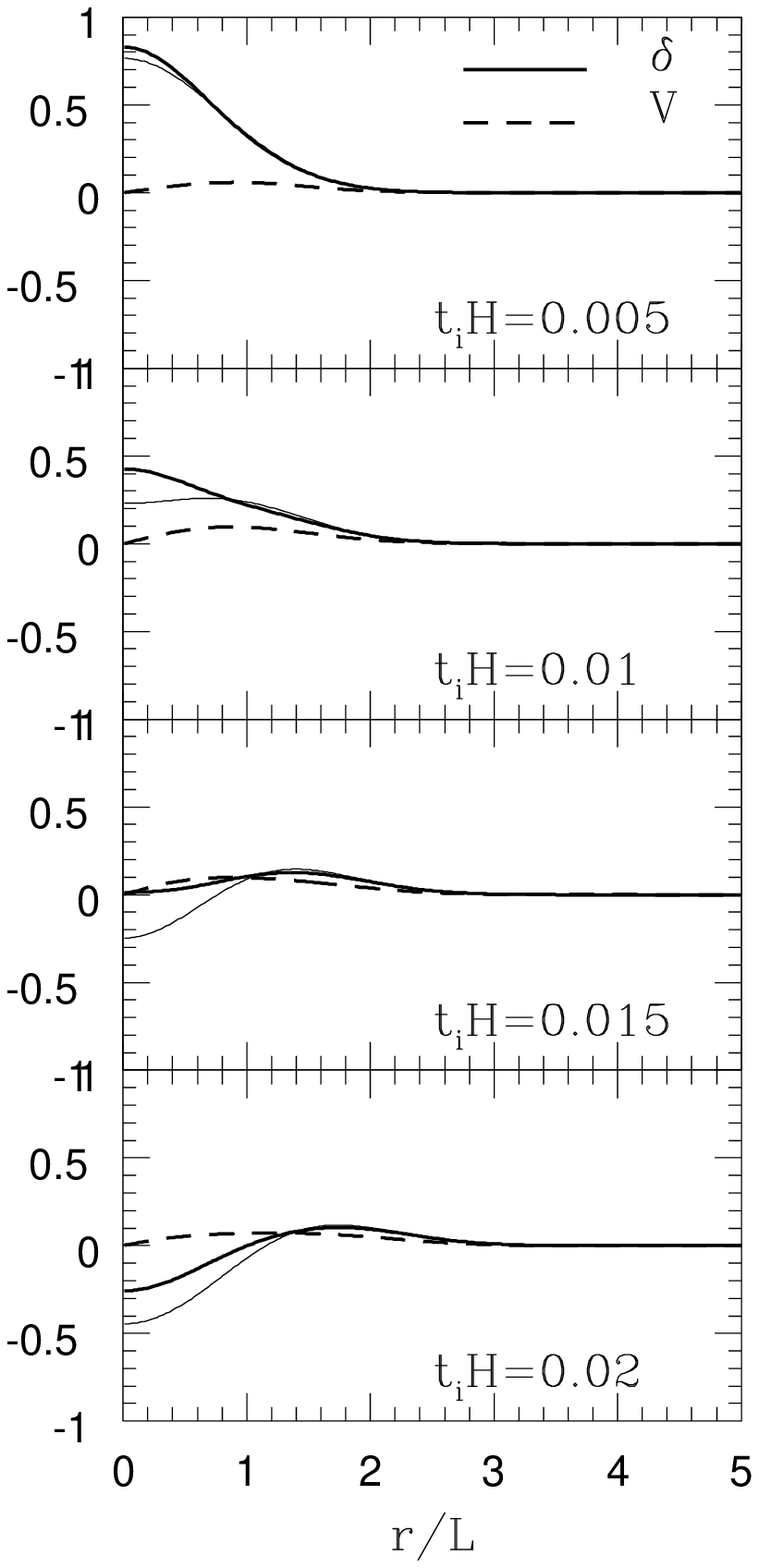}}
\caption{
Time evolution of the density contrast relative to Figure~\ref{fig:w03A03ETLT} (left panels) and Figure~\ref{fig:w03NLETLT} (right panels).
In the left panels the initial amplitude is $A=0.3$, while in the right panels the initial amplitude is $A=1$.
\label{fig:w03L001timeevolution}}
\end{center}
\end{figure}

The validity of our numerical solutions is tested using the following two methods.
First, we use the remaining extra equation (\ref{gmg}), which can be rewritten as
\begin{eqnarray}
&&{e^{-2{\cal E}}\over a^2}\biggl[-2{a'^2\over a^2}-{1\over r}\left({a'\over a}(
1+r{\cal E}'+2r\nu')-{ra''\over a}\right)-{1\over r^2}\left(1+r\nu'-r^2\nu'^2+r{\cal E}'
(1+r\nu')-r^2\nu''\right)
\nonumber
\\
&&~~~~~~~~~~~+{e^{2{\cal E}} \over r^2}\biggr]
+e^{-2\nu}\left[-3{\dot a\over a}\dot {\cal E}-\dot {\cal E}^2+\dot{\cal E} \dot\nu-\ddot{\cal E}
\right]=0.
\label{tracefree}
\end{eqnarray}
The terms in the first bracket of this equation do not include differentiation
with respect to $t$,
while the terms in the second bracket only include differentiation with respect
to $t$. Therefore, we use \eqref{tracefree} to check our numerical calculations.
Figure~\ref{fig:gmg} exemplifies a snapshot of our numerical results, 
demonstrating the balance between the first term (solid curve) and
the second term (dashed curve).
Second, in Appendix~\ref{perturb} we develop the perturbation theory for the 
Lema\^{i}tre model, whose prediction for the density perturbation is plotted 
(thin solid curves) against the fully nonlinear solutions in the top panels 
of Figures \ref{fig:w03L1ETLT}, \ref{fig:w03LNETLTIC}, \ref{fig:w03A03ETLT}, \ref{fig:w03NLETLT}
and in the panels of Figures \ref{fig:w03L1timeevolution} and \ref{fig:w03L001timeevolution}.
When the amplitude of the density contrast is not large ($A\simlt0.1$), 
the perturbation theory well reproduces the full numerical results based on 
the equations presented in Section~\ref{model}.

Figure \ref{fig:w03L1ETLT} shows our numerical results for the case of 
a relativistic fluid ($w=1/3$) with initial parameters
$A=0.3$ and $\ell=LH_i=1$, that is, a characteristic scale of one initial Hubble radius.
The initial spatial curvature is set to zero as explained in Section~\ref{zero}.
The left and right panels of Figure \ref{fig:w03L1ETLT} show 
snapshots at $H_it=5~(\bar a=3.3)$ and 
at $H_it=20~(\bar a=6.4)$, respectively.
In the top panels, the thick solid curve is the density contrast $\delta(t,r)$
and the dashed curve is $V(r)$ defined by equation (\ref{defV}), which is understood as the 
velocity field in the Newtonian description.
One can see that the central overdensity 
propagates away, then a wave-like feature appears at a later time
(see the left panels of Figure~\ref{fig:w03L1timeevolution} for a more detailed time evolution).
This is a consequence of the adopted initial conditions which are 
dominated by decaying modes and cause the density contrast to decrease.
The linearized theory (thin solid curve, see Appendix~\ref{perturb}) is in agreement with the exact solution as the density contrast quickly becomes linear.
In the second panels from the top, the solid curve is the lapse function $\nu(r)$ and the dashed curve is ${\cal E}(r)$. 
The details of the lower panels are explained in the next Section.

In Figure~\ref{fig:w03LNETLTIC} we repeat the calculation for 
the case of uniform initial expansion (see Section~\ref{uniform}).
All the other model parameters are unchanged. 
From the top panels of Figure~\ref{fig:w03LNETLTIC}, one sees that 
in this case the density contrast does not decrease as compared with Figure \ref{fig:w03L1ETLT}
(see also the right panels of Figure~\ref{fig:w03L1timeevolution}).
This should be due to the fact that the initial conditions are dominated by growing modes 
and the characteristic scale of the inhomogeneity is initially larger than the sound horizon scale.
Indeed, from \eqref{shor} one obtains $r_s=L$ at $H_it=1$, $r_s=1.9L$ at $H_it=5$ 
and $r_s=3.7L$ at $H_it=20$.
The increase of $r_s$ at later times explains why the density contrast did not continue to grow but after 
reaching a maximum value of $0.7$ at $H_it=3$ it then decreases to about $0.3$ at $H_it=20$
(see also the right panels of Figure~\ref{fig:w03L1timeevolution}).
The nonlinear central density contrast causes the deviation between the full numerical result and the linear perturbation theory as shown by the right panels of Figure~\ref{fig:w03L1timeevolution}.

Figure~\ref{fig:w03A03ETLT} shows the evolution relative to the initial parameters $A=0.3$ and 
$\ell=H_iL=0.01$ at the times  $H_it=0.005$ (left panels) and $H_it=0.02$ (right panels), respectively.
Figure~\ref{fig:w03NLETLT} shows the evolution relative to the same parameters of Figure 
\ref{fig:w03A03ETLT} except for a larger initial contrast of $A=1$.
The sound horizon scale is in both cases $r_s=58L$ and $r_s=59L$ at $H_it=0.005$ and $H_it=0.02$, 
respectively, much larger than the inhomogeneity scale.
We repeated the computation relative to Figures~\ref{fig:w03A03ETLT}-\ref{fig:w03NLETLT} for 
the case of the uniform initial expansion described in Section~\ref{uniform} and found that 
the evolution is basically unchanged. Therefore, we conclude that the dynamical evolution is 
insensitive to the initial conditions for the curvature function $E(r)$ when the characteristic 
inhomogeneity scale is much smaller than the sound horizon scale.
Equivalently, pressure gradients quickly overcome the initial velocity profile.
By comparing the thick solid curve and the thin solid curve in the top panels
of Figure \ref{fig:w03NLETLT},  we see that the linear 
perturbation description is good at an earlier stage of the time evolution 
even for the case of a nonlinear initial overdensity contrast $A=1$ 
(see also Figure~\ref{fig:w03L001timeevolution}).
However, the discrepancy between linear and nonlinear dynamics becomes large at a later time, when
wave-like features appear. The amplitude of the wave-like 
features disappear faster for the cases of $A\simgt1$ than for the cases of $A\simlt1$, 
which is understood as a nonlinear effect of strong gravity (compare the relative amplitudes in Figure~\ref{fig:w03L001timeevolution}).
Summarizing, the linearized theory agrees better with the exact solution when the contrast is smaller -- as is clear from Figures~\ref{fig:w03A03ETLT}-\ref{fig:w03L001timeevolution} -- and fails in the central region when 
the density contrast is large.

\section{Newtonian potentials from coordinate transformation}
\label{Newton}
We now consider whether the Lema\^{i}tre model can be brought to the form of the perturbed FLRW universe in the Newtonian gauge. 
This problem was considered for the case of a dust fluid ($w=0$) in~\cite{Biswas:2007gi,Paranjape:2008ai,VanAcoleyen:2008cy}.
To our knowledge, the case $w\neq0$ has not been investigated so far.

We consider the coordinate transformation from (\ref{GTBGTB}) --
which we use to obtain the dynamics -- to the perturbed FLRW line element:
\begin{eqnarray}
ds^2=-(1+{2 \psi({\tilde t},{\tilde r})})d{\tilde t}^2+{\bar a}^2({\tilde t})(1+{2\phi({\tilde t},{\tilde r})})
(d{\tilde r}^2+\tilde r^2 d\Omega^2). 
\label{cnp}
\end{eqnarray}
We adopt the following ansatz for the coordinate transformation~\cite[see][]{Paranjape:2008ai}: 
\begin{eqnarray}
{\tilde t}&=&t+\xi^0(t,r),
\label{tttt}
\\
{\tilde r}&=&{a(t,r)r\over {\bar a}(t)}{1\over 1-\xi(t,r)},
\label{trtr}
\end{eqnarray}
which yields:
\begin{eqnarray}
{\partial {\tilde t}\over \partial t}&=&1+\dot \xi^0(t,r) \,,
~~~~~~{\partial {\tilde t}\over \partial r}=\xi^0{}'(t,r) \,,
\nonumber
\\
{\partial {\tilde r}\over \partial t}&=&\left({\dot a(t,r)\over {\bar a}(t)}-{\dot {{\bar a}}(t)\over {{\bar a}}^2(t)}a(t,r)\right){r\over 1-\xi(t,r)}+{a(t,r)r\over {\bar a}(t)}{\dot \xi(t,r)\over (1-\xi(t,r))^2} \,,
\nonumber
\\
{\partial {\tilde r}\over \partial r}&=&{a(t,r)+ra'(t,r)\over {\bar a}(t)}{1\over 1-\xi(t,r)}+{a(t,r)r\over {\bar a}(t)}{\xi'(t,r)\over (1-\xi(t,r))^2} \,.
\nonumber
\end{eqnarray}
From the components $(t,t)$,~$(t,r)$,~$(r,r)$, and $(\theta,\theta)$ of the transformation of the metric, 
$g_{\mu\nu}(x)= \tilde g_{\alpha\beta}(\tilde x) 
{\partial \tilde x^{\alpha}\over \partial x^\mu}
{\partial \tilde x^{\beta}\over \partial x^\nu}$, we have the following equations:
\begin{eqnarray}
&& -e^{2\nu(t,r)}=-(1+2\psi({\tilde t},{\tilde r}))\left\{1+\dot\xi^0(r,t)\right\}^2+
{\bar a}^2({\tilde t})(1+2\phi({\tilde t},{\tilde r}))r^2
\nonumber\\
&&~~~~~~~~~~~~~~~\times
\biggl\{\left({\dot a(t,r)\over {\bar a}(t)}-{\dot {{\bar a}}(t)\over {{\bar a}}^2(t)}a(t,r)\right){1\over 1-\xi(t,r)}+{a(t,r)\over {\bar a}(t)}{\dot \xi(t,r)\over (1-\xi(t,r))^2}\biggr\}^2,
\label{Aeqd}
\\
&&0=-(1+2\psi({\tilde t},{\tilde r}))(1+\dot \xi^0(t,r))\xi^0{}'(t,r)
+{\bar a}^2({\tilde t})(1+2\phi({\tilde t},{\tilde r}))
\nonumber\\
&&~~~~~~~~~~~~~~~\times
\biggl\{\left({\dot a(t,r)\over {\bar a}(t)}-{\dot {{\bar a}}(t)\over {{\bar a}}^2(t)}a(t,r)\right){r\over 1-\xi(t,r)}+{a(t,r)r\over {\bar a}(t)}{\dot \xi(t,r)\over (1-\xi(t,r))^2}\biggr\}
\nonumber\\
&&~~~~~~~~~~~~~~~\times\biggl\{
{a(t,r)+ra'(t,r)\over {\bar a}(t)}{1\over 1-\xi(t,r)}+{a(t,r)r\over {\bar a}(t)}
{\xi'(t,r)\over (1-\xi(t,r))^2}\biggr\},
\label{Aeqc}
\\
&&
a^2(t,r)e^{2{\cal E}(t,r)}=-(1+2\psi({\tilde t},{\tilde r}))(\xi^0{}'(t,r))^2+{\bar a}^2({\tilde t})(1+2\phi({\tilde t},{\tilde r}))
\nonumber\\
&&~~~~~~~~~~~~~~~\times\biggl\{
{a(t,r)+ra'(t,r)\over {\bar a}(t)}{1\over 1-\xi(t,r)}+{a(t,r)r\over {\bar a}(t)}
{\xi'(t,r)\over (1-\xi(t,r))^2}\biggr\}^2,
\label{Aeqb}
\\
&& a^2(t,r)r^2={\bar a}^2(\tilde t)(1+2\phi({\tilde t},{\tilde r})){\tilde r}^2 .
\label{Aeqa}
\end{eqnarray}
Using (\ref{trtr}) and (\ref{Aeqa}), we have 
\begin{eqnarray}
{\bar a}^2({\tilde t})(1+2\phi({\tilde t},{\tilde r}))={\bar a}^2(t)(1-\xi(t,r))^2,
\label{eqa}
\end{eqnarray}
using which one can write (\ref{Aeqd}-\ref{Aeqb}) as
\begin{eqnarray}
&& -e^{2\nu(t,r)}=-(1+2\psi({\tilde t},{\tilde r}))\left(1+\dot\xi^0\right)^2
 +r^2\left(\dot a-{\dot {\bar a}\over {\bar a} }a+{a}{\dot \xi\over 1-\xi}\right)^2,
\label{eqd}
\\
&& 0=-(1+{2\psi({\tilde t},{\tilde r})})(1+\dot \xi^0)\xi^0{}'
+r\left(\dot a-{\dot{\bar a}\over {\bar a}}a+a{\dot \xi\over 1-\xi}\right)
\left(a+ra'+ra{\xi'\over 1-\xi}\right),
\label{eqc}
\\
&& a^2(t,r)e^{2{\cal E}(t,r)}=-(1+{2\psi({\tilde t},{\tilde r})})\left(\xi^0{}'\right)^2
 +\left(a+ra'+ra{\xi'\over 1-\xi}\right)^2.
\label{eqb}
\end{eqnarray}
The four-velocity of the perfect fluid $v^\mu=(e^{-\nu},0,0,0)$ becomes $\tilde v^\mu=({\tilde v}^{\tilde t}, {\tilde v}^{\tilde r},0,0)$ in the perturbed FLRW metric, where:
\begin{eqnarray}
{\tilde v}^{\tilde t}&=&{\partial {\tilde t} \over \partial t}e^{-\nu}=(1+\dot\xi^0(t,r))e^{-\nu(t,r)},
\\
{\tilde v}^{\tilde r}&=&{\partial {\tilde r} \over \partial t}e^{-\nu}=r\left\{
\left({\dot a(t,r)\over {\bar a}(t)}-{\dot{\bar a}(t)\over {\bar a}^2(t)}a(t,r)\right){1\over 1-\xi(t,r)}
+{a(t,r)\over {\bar a}(t)}{\dot\xi(t,r)\over (1-\xi(t,r))^2}
\right\}e^{-\nu(t,r)},
\nonumber\\
\end{eqnarray}
and we define the peculiar velocity as: 
\begin{eqnarray}
V={{\tilde v}^{\tilde r}\over {\tilde v}^{\tilde t}} \,.
\label{defV}
\end{eqnarray}

The third panels from the tops of Figures \ref{fig:w03L1ETLT}-\ref{fig:w03LNETLTIC} and \ref{fig:w03A03ETLT}-\ref{fig:w03NLETLT} show
the profiles of $\xi(r)$ and $H_i\xi^0(r)$, while the bottom
panels show $\phi$, $\psi$ and $\phi+\psi$. 
In each panel, the full nonlinear transformation defined by equations (\ref{eqa}--\ref{eqb}) is used.

From Figures~\ref{fig:w03L1ETLT} and~\ref{fig:w03LNETLTIC} one can see that for a relativistic fluid ($w=1/3$) with 
initial density contrast $A=0.3$ and characteristic scale $\ell=1$, the relation $\psi+\phi={\cal O}(\phi^2)$ is satisfied.
In Figure~\ref{fig:w03L1ETLT} the potentials are well in the linear regime.
This is explained by the fact that the adopted initial conditions are 
dominated by decaying modes and the density contrast quickly becomes linear.
In Figure~\ref{fig:w03LNETLTIC}, instead, the initial conditions are dominated by growing modes and the density contrast grows. This causes larger values of the potentials which, at the initial times, are only marginally within the linear regime. This is essentially due to the fact that at initial time the inhomogeneity is horizon scale. Even though the amplitude of the potentials becomes mildly nonlinear reaching the numerical value of $0.1\sim0.2$, the Newtonian description works in the sense that $\psi+\phi={\cal O}(\phi^2)$ is satisfied.

Figures~\ref{fig:w03A03ETLT} and~\ref{fig:w03NLETLT} show that in the case of sub-sound horizon perturbations the Newtonian 
potentials satisfy $|\phi|, |\psi|\ll1$, which indicates that the perturbed Newtonian 
description is valid even when 
the density contrast is large (see Figure~\ref{fig:w03NLETLT})).
However, the relation $\psi+\phi={\cal O}(\phi^2)$ is not satisfied, that is, there is an effective anisotropic stress, a well known prediction of second-order perturbation theory~\cite[see, e.g.,][]{Bruni:1996im,Matarrese:1997ay,Hu:1998tj,Nakamura:2004wr,Tomita:2005et,Ballesteros:2011cm}.
Interestingly, the same conclusion is reached for a linear initial contrast ($A=0.3$, see Figure~\ref{fig:w03A03ETLT}) as well as for a nonlinear initial contrast ($A=1$, see Figure~\ref{fig:w03NLETLT}), and, as said before, is independent of the adopted initial conditions for the curvature function $E(r)$.
In other words, the Lema\^{i}tre model allows us to study in an exact nonlinear fashion the onset of anisotropic stress in fluids with non-negligible pressure.

\section{Equations in spatially conformally flat coordinates}
\label{conform}

The metric~(\ref{GTBGTB}) uses comoving coordinates. 
It is useful to re-analyze the problem with coordinates which can be directly related to the Newtonian potentials. 
To this end, we now consider the following metric:
\begin{eqnarray}
ds^2=-e^{2\Psi(t,r)}dt^2+{\bar a}^2(t)e^{2\Phi(t,r)} 
(d r^2+ r^2d\Omega^2) \,.
\end{eqnarray}
The four-velocity of the fluid is now $u^\alpha=\gamma(e^{-\Psi}, \bar a^{-1}e^{-\Phi} V_p,0,0)$
where $\gamma=(1-V_p^2)^{-1/2}$. 
Einstein's equation leads then to:
\begin{eqnarray}
G^t{}_t&=&{e^{-2\Phi}\over {\bar a}^2}\left[\Phi'{}^2+2\Phi''+4{\Phi'\over r}\right]
-3 e^{-2\Psi} (H+\dot\Phi)^2
=-8\pi G\left(
(\rho+P)\gamma^2-P\right),
\label{Gttb}
\\
G^t{}_r&=&-2e^{-2\Psi}\left[(H+\dot\Phi)\Psi'-\dot\Phi'\right]
=8\pi G{\bar a} e^{\Phi-\Psi}(\rho+P)V_p\gamma^2,
\\
G^r{}_r&=&{e^{-2\Phi}\over r{\bar a}^2}\left[2\Psi'(1+r\Phi')+\Phi'(2+r\Phi')\right]
+e^{-2\Psi}\left[-H^2-2{\ddot {\bar a}\over {\bar a}}\right.\nonumber
\\&&~~~~~~~~
\left.
+2H(\dot\Psi-3\dot\Phi)+2\dot\Psi\dot\Phi-3\dot\Phi^2-2\ddot\Phi\right]
=8\pi G ( (\rho+P) V_p^2 \gamma^2+P),
\\
G^\theta{}_\theta&=&
{e^{-2\Phi}\over r{\bar a}^2}\left[\Phi'+\Psi'+r(\Phi''+\Psi'')+r\Psi'{}^2\right]
+e^{-2\Psi}\left[-H^2-2{\ddot {\bar a}\over {\bar a}}\right.\nonumber
\\&&~~~~~~~~
\left.
+2H(\dot\Psi-3\dot\Phi)+2\dot\Psi\dot\Phi-3\dot\Phi^2-2\ddot\Phi\right]
=8\pi GP,
\end{eqnarray}
where, as before, $H=\dot {\bar a}/{\bar a}$. Finally, the fluid equations are:
\begin{eqnarray}
&&{\bar a}(t) e^{\Phi}\biggl\{\dot \rho-V_p^2\dot P+(\rho+P)(3-V_p^2)(H+\dot\Phi)\biggr\}
\nonumber
\\
&&~
+e^{\Psi}\biggl\{V_p(\rho'-P')+(\rho+P)(V_p'+{2\over r}V_p+2V_p\Phi')\biggr\}
=0,
\label{dotrhob}
\\
&&{\bar a}(t) e^{\Phi}\biggl[(\rho+P)(1+V_p^2)\dot V_p+V_p(1-V_p^2)\biggl\{
\dot\rho+\dot P+4(\rho+P)(H+\dot\Phi)\biggr\}\biggr]
\nonumber
\\
&&~+e^{\Psi}\biggl[(1-V_p^2)\biggl\{P'+V_p^2\rho'+(\rho+P)\Bigl(2V_p^2({1\over r}+\Phi')+(1+V_p^2)\Psi'\Bigr)\biggr\}+2(\rho+P)V_pV_p'\biggr]=0.
\nonumber
\\
\label{dotvpb}
\end{eqnarray}

\subsection{First-order equations}

Then we expand perturbatively the dynamical variables of the previous equations:
\begin{eqnarray}
\Phi(t,r)&=&{\Phi_1}(t,r)+{\Phi_2}(t,r)+\cdots,
\\
\Psi(t,r)&=&{\Psi_1}(t,r)+{\Psi_2}(t,r)+\cdots,
\\
\rho(t,r)&=&{\bar \rho}(t)+\delta\rho_1(t,r)+\delta\rho_2(t,r)+\cdots,
\\
P(t,r)&=&{\bar P}(t)+\delta P_1(t,r)+\delta P_2(t,r)+\cdots,
\\
V_p(t,r)&=&V_1(t,r)+V_2(t,r)+\cdots,
\end{eqnarray}
so that, at the first order, the gravitational field equations become:
\begin{eqnarray}
&&{1\over {\bar a}^2}\left(\Phi_1''+{2\over r}\Phi_1'\right)+3H^2\Psi_1-3H\dot\Phi_1=-4\pi 
G\delta\rho_1,
\label{CFa}
\\
&&-H\Psi_1'+\dot\Phi_1'=4\pi G {\bar a} ({\bar \rho}+{\bar P}) V_1,
\label{CFb}
\\
&&\Psi_1''+\Phi_1''-{1\over r}(\Psi_1'+\Phi_1')=0 .
\label{CFc}
\end{eqnarray}
Similarly, the first-order fluid equations are:
\begin{eqnarray}
&&\dot \Delta_1 +{1+w\over a}\left(V_1'+{2\over r}V_1+3{\bar a}\dot\Phi_1\right)
+3H(c_s^2-w)\Delta_1=0,
\label{CFd}
\\
&&(1+w)\dot V_1+c_s^2{\Delta'_1\over {\bar a}}+(1+w){\Psi_1'\over {\bar a}}+H(1+w)(1-3w)V_1=0,
\label{CFe}
\end{eqnarray}
$\Delta_1= \delta \rho_1/\bar \rho$, $\bar P=w\bar \rho$ and $\delta P_1=c_s^2\delta\rho_1$, and we have assumed the fluid equation of state parameter $w$ and sound speed $c_s$ to be constant.

We then Fourier expand the first-order quantities:
\begin{eqnarray}
\Phi_1(t,r)&=&\widetilde \Phi(k,t) j_0(kr),
\\
\Psi_1(t,r)&=&\widetilde \Psi(k,t) j_0(kr),
\\
\Delta_1(t,r)&=&\widetilde \Delta(k,t) j_0(kr),
\\
V_1(t,r)&=&\widetilde V(k,t) j_1(kr),
\end{eqnarray}
where $j_\ell(z)$ is the $\ell$-th order spherical Bessel function of the first kind, so that 
equations (\ref{CFa}-\ref{CFe}) become:
\begin{eqnarray}
&&-k^2{ \widetilde \Phi}+3{\cal H}^2\widetilde\Psi-3{\cal H}{\widetilde \Phi}_{,\eta}
=-{3\over 2}{\cal H}^2\widetilde\Delta,
\label{perter1}
\\
&&k({\cal H}\widetilde \Psi-{\widetilde\Phi}_{,\eta})={3\over 2}(1+w){\cal H}^2\widetilde V,
\label{perter2}
\\
&&\widetilde\Psi+\widetilde\Phi=0,
\label{perter3}
\\
&&{\widetilde \Delta}_{,\eta} +{(1+w)}\left(k\widetilde V+3{\widetilde\Phi}_{,\eta}\right)+3{\cal H}(c_s^2-w)\Delta=0,
\label{perter4}
\\
&&{\widetilde V}_{,\eta}-{c_s^2k\over 1+w}{\widetilde\Delta}-k{\widetilde\Psi}+{\cal H}(1-3w)\widetilde 
V=0,
\label{perter5}
\end{eqnarray}
where $,\eta$ denotes  differentiation 
with respect to the conformal time defined by $d\eta=dt/{\bar a}$, ${\cal H}={\bar a}_{,\eta}/{\bar a}$ and we used the background Friedmann equation ${\cal H}^2=8\pi G\bar a^2{\bar \rho}/3$.
In particular, as ${\bar \rho}\propto a^{-3(1+w)}$, the Friedmann equation yields ${\bar a}(t)\propto t^{2/3(1+w)}\propto \eta^{2/(1+3w)}$.
Note that it is $\widetilde\Psi+\widetilde\Phi=0$ at the first order of 
perturbative expansion.

By solving equations (\ref{perter1}) and (\ref{perter2}) for $\widetilde\Delta$ and $\widetilde V$, respectively, equation (\ref{perter5}) gives:
\begin{eqnarray}
{\partial^2 \widetilde\Phi \over \partial \eta^2}+{6(1+c_s^2)\over (1+3w)\eta}
{\partial   \widetilde\Phi \over \partial \eta}+\biggl(c_s^2k^2+{12(c_s^2-w)\over (1+3w)^2}\biggr)\widetilde\Phi=0,
\end{eqnarray}
where we used $\widetilde\Phi+\widetilde\Psi=0$. The solution is given by:
\begin{eqnarray}
\widetilde\Phi=\alpha_k \eta^{-\mu_1}J_{\mu_2}(c_sk\eta)
+\beta_k\eta^{-\mu_1}N_{\mu_2}(c_sk\eta),
\label{SolutionC}
\end{eqnarray}
where we defined
\begin{eqnarray}
  \mu_1={5+6c_s^2-3w\over 2(1+3w)}, 
~~~~~\mu_2={\sqrt{25+12c_s^2+36c_s^4+18w-36c_s^2 w+9w^2}\over 2(1+3w)},
\end{eqnarray}
and $J_\nu(z)$ and $N_\nu(z)$ are the 
Bessel functions of the first and second kind, respectively, and $\alpha_k$ 
and $\beta_k$ are the coefficients specified by the initial 
conditions.
In the limit $w=c_s^2$, it is 
\begin{eqnarray}
\mu_1=\mu_2={5+3w\over2(1+3w)}\equiv\mu,
\end{eqnarray}
which is consistent with the results of Ref.~\cite{Baumann:2007zm}.

\subsection{Second-order equations}

Next, we consider the equations for the second-order perturbations. 
From the second-order Einstein's equation, we have
${\delta}G^{\theta(2)}_{~\theta}-{\delta}G^{r(2)}_{~r}=8\pi G({\delta}T^{\theta(2)}_{~\theta}-{\delta}T^{r(2)}_{~r})$, 
which gives: 
\begin{eqnarray}
&&
-{\Phi_1}'{}^2-{1\over r}{\Phi_2}'-2{\Phi_1}'{\Psi_1}'+{\Psi_1}'{}^2-{1\over r}{\Psi_2}'+{\Phi_2}''+ {\Psi_2}''\nonumber\\
&&~~~~~~+{2}
{\Phi_1}\bigl({1\over r}({\Phi_1}'+{\Psi_1}')-{\Phi_1}''-{\Psi_1}''\bigr)
=-8\pi G (1+w){\bar a}^2{\bar \rho} V_1^2.
\label{counter}
\end{eqnarray}
Since we aim at understanding the cause behind the violation of $\phi+\psi \simeq 0$ (see the results of Section~\ref{Newton}), we now focus on the relation between $\Phi_2$ and $\Psi_2$. 
Using the first-order equation ${\Phi_1}+{\Psi_1}=0$, (\ref{counter}) yields
\begin{eqnarray}
{\Phi_2}''+{\Psi_2}''-{1\over r}({\Phi_2}'+{\Psi_2}')=-2(\Phi_1'^2+4\pi G(1+w)
{\bar \rho}(t) {\bar a}^2V_1^2),
\label{bracket2}
\end{eqnarray}
which can be integrated in order to make explicit the source of $\Phi+\Psi$:\footnote{
Note that we may write $\phi+\psi=\Phi_2+\Psi_2+2\Phi_1^2$ up to the second order of 
perturbations for the potentials defined in section 4.}
\begin{eqnarray}
{\Phi_2}(t,r)+{\Psi_2}(t,r)=-2\int_r^\infty dz z\int_z^\infty ds{1\over s}\biggl[
\Phi_1'^2(t,s)+4\pi G(1+w){\bar \rho}(t) {\bar a}^2V_1^2(t,s)
\biggr].
\label{bracket}
\end{eqnarray}
Next we compare the two source terms inside the square brackets of (\ref{bracket}). 
Using the first-order solution, the first term is estimated as
\begin{eqnarray}
&&\Phi_1'^2\sim k^2\widetilde\Phi^2 \,.
\end{eqnarray}
Regarding the second term, from (\ref{perter2}) one has (${\widetilde\Phi}_{,\eta}=c_s k \widetilde\Phi$):
\begin{eqnarray}
&&\widetilde V\sim-{2k({\cal H}+c_sk)\widetilde \Phi\over 3(1+w){\cal H}^2},
\label{Vtilap}
\end{eqnarray}
which allows us to conclude that:
\begin{eqnarray}
&&4\pi G(1+w){\bar \rho}(t) {\bar a}^2
V_1^2 \sim{2k^2\over 3(1+w)}
\left(1+{c_sk\over {\cal H}}\right)^2
\widetilde\Phi^2.
\label{secter}
\end{eqnarray}
Therefore, the second term inside the square brackets of (\ref{bracket}) is larger than the first term if:
\begin{eqnarray}
&&{2\over3(1+w)}\left({1}+{c_sk\over {\cal H}}\right)^2\gg1 \,,
\label{condition}
\end{eqnarray}
which simplifies to ${c_s k/ {\cal H}}\gg{\cal O}(1)$, as long as $w$ is not close to $-1$.
Thus the relation  $\Phi_2+\Psi_2\sim {\cal O}(\Phi_1^2)$ may become invalid when the inhomogeneity scale is smaller than 
the sound horizon and the contribution of the
velocity field is significant.

We conclude the analysis by carefully estimating the value of $\Phi_2+\Psi_2$.
Introducing $L$ as the typical
scale of the inhomogeneity, from \eqref{bracket2} one finds that the first term on the right-hand side of (\ref{bracket}) is of order $\Phi_1^2$.
For scales smaller than the sound horizon ($L\ll c_s{\cal H}^{-1}$) from \eqref{Vtilap} one has
$V_1\sim c_s{\Phi_1} /(1+w)(L{\cal H})^2$.
Consequently, in this limit, the second term on the right-hand side of (\ref{bracket}) is of the order $c_s^2{\Phi_1}^2/(1+w)(L{\cal H})^2$.
If we further assume that $L\ll {\cal H}^{-1}$, that is a sub-horizon inhomogeneity, from (\ref{perter1}) one obtains for the density contrast 
$\Delta_1\sim{\Phi_1}/(L{\cal H})^2$. 
Then, the second term on the right-hand side of (\ref{bracket}) becomes $c_s^2\Delta_1{\Phi_1}/(1+w)$.
We can finally conclude that:
\begin{eqnarray}
\Phi_2+\Psi_2\sim {\rm max}[{\cal O}({\Phi_1}^2),{\cal O}(c_s^2{\Phi_1}\Delta_1)]
\label{fres}
\end{eqnarray}
for an inhomogeneity whose characteristic scale is smaller than both the Hubble horizon and the sound horizon, so long as $w$ is not close to $-1$.
The result of \eqref{fres} is based on a perturbative expansion. Therefore, the 
second-order quantities must be smaller than the first-order 
ones, that is, $|\Phi_2|, |\Psi_2| < |\Phi_1|, |\Psi_1|$.
From \eqref{fres} one then concludes that $c_s^2{\Phi_1}\Delta_1 < \Phi_1$, so that for consistency of the perturbative expansion we require $\Delta_1<1$.

On the other hand, if the characteristic scale of the inhomogeneity is larger 
than the sound horizon ($L\gg c_s{\cal H}^{-1}$), the second term on the right-hand side 
of (\ref{bracket}) becomes ${\Phi_1}^2/(1+w)$ (see \eqref{secter}), which is of the same
order as the first term as long as $w$ is not close to $-1$.
The analytical results of this Section confirm what we have found numerically in the analysis relative to 
Figures \ref{fig:w03L1ETLT}-\ref{fig:w03LNETLTIC} and \ref{fig:w03A03ETLT}-\ref{fig:w03NLETLT}.

\section{Conclusions}
\label{conclusions}

We studied the Lema\^{i}tre model with non-negligible pressure in a 
cosmological setup using a numerical method. We investigated the 
validity of the Newtonian description of the Lema\^{i}tre model by transforming its metric into the perturbed FLRW metric. 
For the case of a relativistic fluid ($w=1/3$), the Newtonian 
description is valid as the amplitudes 
of the potentials remain small $|\phi|,|\psi|\ll1$, even when the density 
contrast is of the order ${\cal O}(1)$, as long as the inhomogeneity 
is sub horizon. However, the relation $\phi+\psi={\cal O}(\phi^2)$, which 
holds in the linear cosmological perturbation theory, ceases to be valid. 
In the dust-dominated case, it is known that the deviation is 
expressed by the second-order potential, $\phi+\psi={\cal O}(\phi^2)$. 
However, this is not
the case for an inhomogeneous fluid with non-negligible pressure.  
This suggests that density inhomogeneities in a fluid with
non-negligible pressure may effectively give rise to anisotropic stress.
Therefore, the Lema\^{i}tre model gives us the chance to study in an exact nonlinear 
fashion the generation of anisotropic stress in fluids with non-negligible pressure, 
thus extending previous work based on second-order perturbation theory~\cite[see, e.g.,][]{Bruni:1996im,Matarrese:1997ay,Hu:1998tj,Nakamura:2004wr,Tomita:2005et,Ballesteros:2011cm}.
From the analysis of the second-order perturbations, we found that 
$\phi+\psi={\cal O}(\phi^2)$ is not valid when the characteristic scale of 
the inhomogeneity is smaller than the sound horizon, in which case it is
$\phi+\psi={\rm max}[{\cal O}(\phi^2),{\cal O}(c_s^2\phi\delta)]$, 
where $\delta$ is the density contrast, as long as $w$ is not close to $-1$.
As this estimation is based on a perturbative approach, $\delta<1$ is required for consistency.
We also conclude that 
$\psi+\phi={\cal O}(\phi^2)$ holds in general when the characteristic 
scale of the inhomogeneities is larger than the sound horizon scale, 
unless the amplitude of the inhomogeneities is nonlinear and 
$w$ is close to $-1$.

\section*{Acknowledgment}
Valerio Marra is supported by the Brazilian research agency CNPq.
Viatcheslav Mukhanov is supported by TRR 33 ``The Dark Universe'' and the Cluster of
Excellence EXC 153 ``Origin and Structure of the Universe''. 
This work was supported by MEXT/JSPS KAKENHI Grant Numbers 15H05895 and 15H05888.

\appendix

\section{Perturbative approach for the Lema\^{i}tre model}
\label{perturb}

Here, we develop the perturbation theory for the Lema\^{i}tre model.
We adopt the following ansatz for the variables, assuming the perturbative 
quantities are of an order less than unity:
\begin{eqnarray}
\rho&=&{\bar \rho}(t)(1+\delta_1+\delta_2+\cdots),
\\
a&=&{\bar a}(t)(1+\zeta_1+\zeta_2+\cdots),
\\
\nu&=&\nu_1+\nu_2+\cdots,
\\
{\cal E}&=&{\cal E}_i(r)+{\cal E}_1(t,r)+{\cal E}_2(t,r)+\cdots .
\end{eqnarray}
Note that we consider the case where $\nu_0(t)=0$ and that
${\cal E}_i(r)$ is included as a first-order perturbation. 
Up to the first order of the perturbative expansions,  
(\ref{mtrr2}) and  (\ref{mtrt2}) become: 
\begin{eqnarray}
M'(t,r)&=&4\pi Gr^2{\bar \rho}{\bar a}^3(1+\delta_1+3\zeta_1+r\zeta_1') \,,
\label{pmtrr2}
\\
\dot M(t,r)&=&-4\pi Gr^3w{\bar \rho}{\bar a}^3({\bar H}+{\bar H}(\delta_1+3\zeta_1)+\dot\zeta_1) \,,
\label{pmtrt2}
\end{eqnarray}
where:
\begin{eqnarray}
&&M(t,r)={1\over 2}r{\bar a}\Bigl(r^2{\bar a}^2{\bar H}^2\Bigr)
+{1\over 2}r{\bar a}\biggl[2{\cal E}_i+2{\cal E}_1-2r\zeta_1'+r^2{\bar a}^2{\bar H}(3H\zeta_1
-2{\bar H}\nu_1+2\dot\zeta_1)\biggr].
\end{eqnarray}
From equation (\ref{pmtrr2}), we have
\begin{eqnarray}
&&
{\cal E}_i+{\cal E}_1+r({\cal E}_i'+{\cal E}_1')-r^2\zeta_1''-2r\zeta_1'
=r^2\biggl[{3\over 2}({\bar H}{\bar a})^2\delta_1-{\bar H}{\bar a}^2
(3\dot\zeta_1+r\dot\zeta_1')+({\bar H}{\bar a})^2(3\nu_1+r\nu_1')\biggr].
\nonumber\\
\label{pmtrr2dash}
\end{eqnarray}
Equations (\ref{sigmasol}) and (\ref{cale2}) lead to
\begin{eqnarray}
\nu_1(t,r)&=&-{w\over 1+w}\delta_1,
\label{nuichiap}
\\
{\cal E}_1(t,r)&=&-{1\over 1+w}\biggl\{\delta_1(t,r)-\delta(t_i,r)\biggr\}
-3\biggl\{\zeta_1(t,r)-\zeta_1(t_i,r)\biggr\},
\label{calEap}
\end{eqnarray}
respectively. Equations (\ref{aEMe}) and (\ref{EMfz}) reduce to
\begin{eqnarray}
\dot\delta_1&=&-r wH\delta_1' -(1+w)(3\dot\zeta_1+r\dot\zeta_1') \,,
\label{contapap} \\
\dot{\cal E}_1&=&r\left({wH\over 1+w}\delta_1'+\dot\zeta_1'\right ) \,,
\label{contaxap}
\end{eqnarray}
respectively.
Combining (\ref{pmtrr2dash}), (\ref{nuichiap}), (\ref{contapap}) and (\ref{contaxap}), 
we finally have:
\begin{eqnarray}
&&
{\partial^2\delta_1\over \partial\eta^2}+\biggl [{1\over\cal H}{\partial {\cal H}\over \partial \eta}
+{3{\cal H}\over 2}\Bigl(1-w\Bigr)\biggr ]{\partial\delta_1\over \partial\eta}
+3{\partial {\cal H}\over \partial \eta}\Bigl(1-w\Bigr)\delta_1
-w\biggl(\delta_1''+{2\over r}\delta_1'\biggr)=0.
\nonumber\\
\end{eqnarray}
In the case $w\neq 0$, the solution is:
\begin{eqnarray}
\delta_1(t,r)=\int dk\Bigl [ \tilde A_kD_+(w^{1/2}k\eta)+\tilde B_kD_-(w^{1/2}k\eta)\Bigr ]{\cal U}_k(r),
\end{eqnarray}
where: 
\begin{eqnarray}
&&D_+(z)=z^{2-\mu} J_\mu(z) \,,\\
&&D_-(z)=z^{2-\mu} N_\mu(z) \,,
\end{eqnarray}
with $\mu=(5+3w)/(2+6w)$, and the function ${\cal U}_k(r)$ is defined as:
\begin{eqnarray}
{\cal U}_k(r)=\sqrt{2\over \pi} k j_0(kr),
\end{eqnarray}
which satisfies the relation
$\int_0^\infty dr r^2{\cal U}_k(r){\cal U}_{k'}(r)=\delta(k-k')$.
In the case $w=0$, the growing and the decaying solutions are $D_+\propto \eta^2$
and $D_-\propto \eta^{-3(1+w)/(1+3w)}$, respectively, being independent of the wavenumber $k$.

As discussed in Section~\ref{dynamics}, at the initial time $t=t_i$ we choose $a(t_i,r)=1$, 
that is, $\zeta_1(t_i,r)=0$. 
Therefore, at the initial time $t_i$ from \eqref{pmtrr2dash} we have:
\begin{eqnarray}
{3\over 2}\delta_1-{\bar H}^{-1}(3\dot\zeta_1+r\dot\zeta_1')+3\nu_1+r\nu_1'={1\over (\bar H r)^2}
({\cal E}_i(r)+r{\cal E}_i'(r)) \,,
\end{eqnarray}
which, using (\ref{nuichiap}) and (\ref{contapap}), becomes: 
\begin{eqnarray}
\dot\delta(t_i,r)=-\bigl(1-w\bigr){3{\bar H}(t_i)\over 2}\delta(t_i,r)
+{1+w\over r^2\bar H(t_i)}\bigl({\cal E}_i(r)+r{\cal E}_i'(r)\bigr).
\end{eqnarray}
The Fourier coefficients $\tilde A_k$ and $\tilde B_k$ are then given by:
\begin{eqnarray}
&&\tilde A_k={1\over W}\biggl(+C_k\dot D_-(z)+
\biggl[C_k(1-w){3\over 2}{\bar H}(t)-{1+w\over \bar H(t)}F_k\biggr]
D_-(z)\biggr)\bigg|_{t=t_i},
\\
&&\tilde B_k={1\over W}\biggl(-C_k\dot D_+(z)-\biggl[C_k(1-w){3\over 2}{\bar H}(t)
-{1+w\over \bar H(t)}F_k\biggr]
D_+(z)\biggr)\bigg|_{t=t_i},
\end{eqnarray}
where we defined 
\begin{eqnarray}
&&W=(D_+(z)\dot D_-(z)-\dot D_+(z) D_-(z))={2\over \pi}
(w^{1/2}k\eta)^{3-2\mu}{w^{1/2}k\over a},
\\
&&C_k=\int_0^\infty dr r^2 \delta(t_i,r) {\cal U}_k(r),
\\
&&F_k=\int_0^\infty dr {({\cal E}_i+r{\cal E}_i')} {\cal U}_k(r).
\end{eqnarray}
The following formula is useful to evaluate $C_k$ as we adopt $\delta_i(r)=A e^{-r^2/L^2}$ (see Section~\ref{dynamics}):
\begin{eqnarray}
\int_0^\infty dr r^2 e^{-r^2/L^2} {\cal U}_k(r)={kL^3\over 2^{3/2}}e^{-k^2L^2/4}. 
\end{eqnarray}
As $\zeta_1(t_i,r)=0$, the solution for $\zeta_1$ is:
\begin{eqnarray}
\zeta_1(t,r)={1\over r^3}\int_r^\infty dyy^2 \int_{t_i}^t ds{y w{\bar H} \delta_1'(y,s)+\dot \delta_1(y,s)
\over 1+w} \,.
\end{eqnarray}
Then ${\cal E}_1$ and $\nu_1$ are given by (\ref{calEap}) and (\ref{nuichiap}), respectively.

\bibliography{refs}

\end{document}